# Intertwined topological phases in $TaAs_2$ nanowires with giant magnetoresistance and quantum coherent surface transport

Anand Roy[a*], Anna Eyal[b], Roni Majlin Skiff,[c] Barun Barick[d], Samuel D. Escribano[d], Olga Brontvein[e], Katya Rechav[e], Ora Bitton[e], Roni Ilan[c] and Ernesto Joselevich[a*]

[a]Department of Molecular Chemistry and Materials Science, Weizmann Institute of Science, Rehovot 76100, Israel; [b]Physics department, Technion, Haifa, 32000, Israel; [c]Raymond and Beverly Sackler School of Physics and Astronomy, Tel Aviv 69978, Israel [d]Department of Condensed Matter Physics, Weizmann Institute of Science, Rehovot 76100, Israel. [e]Chemical Research Support, Weizmann Institute of Science, Rehovot 76100, Israel. e-mail: ernesto.joselevich@weizmann.ac.il, anand-kumar.roy@weizmann.ac.il.

Nanowires (NWs) of topological materials are emerging as an exciting platform to probe and engineer new quantum phenomena that are hard to access in bulk phase. Their quasi-one-dimensional geometry and large surface-to-bulk ratio unlock new expressions of topology and highlight surface states. $TaAs_2$, a compensated semimetal, is a topologically rich material harboring nodal-line, weak topological insulator (WTI), $C_2$-protected topological crystalline insulator, and Zeeman field-induced Weyl semimetal phases. We report the synthesis of $TaAs_2$ NWs *in situ* encapsulated in a dielectric $SiO_2$ shell, which enabled us to probe rich magnetotransport phenomena, including metal-to-insulator transition and strong signatures of topologically non-trivial transport at remarkably high temperatures, direction-dependent giant positive and negative magnetoresistance, and a double pattern of Aharonov-Bohm oscillations, demonstrating coherent surface transport consistent with the two Dirac cones of a WTI surface. The coexistence and susceptibility of topological phases to external stimuli have potential applications in spintronics and nanoscale quantum technology.

The discovery of new materials with topologically protected electronic bands has been shown to create vast opportunities for fundamental science and technology advances.[1, 2] In topological semimetals (TSMs), electronic band crossings in the Brillouin zone (BZ) near the Fermi energy ($E_f$) give rise to bulk Dirac/Weyl/nodal-line gapless states and distinct topological surface states (TSS).[3] These bulk and surface states are expected to have exotic properties potentially stabilized by different crystal symmetries.[3] Examining and uncovering the interplay between crystal symmetries and band topologies is thus desirable.

In this aspect, the family of transition metal mono- and dipnictides (TX and $TX_2$, respectively; wherein T= V, Nb, Ta, La, and X= P, As, Sb) provides a rich platform.[4] Non-centrosymmetric TX compounds inherit singly degenerate linear band crossings due to the broken inversion symmetry, and possess Weyl cones of opposite chirality in the bulk and Fermi arcs at the surface.[4, 5] In contrast, centrosymmetric $TX_2$ compounds inheriting doubly degenerate bands due to preserved inversion symmetry, offer bulk nodal-line states without spin-orbit coupling (SOC), and a rare quantum state of matter, namely 3D weak-topological-insulator (WTI), when SOC is included.[6] The WTI nature of $TaAs_2$ and other $TX_2$ analogs is predicted based on its topological invariants (0;111).[6, 7] Notably, TSS harboring a pair of Dirac cones appear only on specific crystallographic surfaces in WTIs, unlike strong topological insulators (STIs), where all the surfaces must harbor an odd number of Dirac cones.[8, 9] This anisotropic nature of WTIs enables surface engineering to obtain topologically trivial (gapped) and non-trivial (gapless) surfaces using different orientations of the same crystal.[8] Due to this anisotropy, WTIs are rare and have so far been reported in only a few 1D and 2D van der Waals (vdW) materials, including β-$Bi_4X_4$ (X = Br, I), $ZrTe_5$ and $Bi_2TeI$.[10, 11, 12, 13, 14, 15] To the best of our knowledge, $TX_2$ compounds are the only documented example of inherent 3D WTIs crystals.[6] The richness of band topologies in $TaAs_2$ and other $TX_2$ analogs further stems from the existence of a two-fold rotation $C_2$ (010) symmetry-protected topological crystalline insulator (TCI) state with a pair of type-II Dirac cones on the {010} surfaces.[16] In these rotational symmetry-protected TCIs, Dirac cones associated to the rotation axis can appear not only at high symmetry points but also at generic **k** points in the surface BZ.[17, 18] These quantum states can further host helical modes on hinges, which could provide a unique platform for realizing and manipulating zero-dimensional Majorana zero modes (MZM) in the proximity of a superconductor.[17, 19] Such rich electronic band topologies in $TaAs_2$ and other $TX_2$ analogs are expected to bring intriguing properties.

In 2016, three seminal articles reported a magnetic field-induced metal-to-insulator (MI) phase transition, followed by a resistivity plateau, non-saturating giant magnetoresistance (GMR ~ $10^5$ order) with Shubnikov-de Hass (SdH) oscillations, and non-trivial Berry phase in bulk single-crystals of $TaAs_2$.[20, 21, 22] The same year, large negative magnetoresistance (MR) in bulk $TaAs_2$ was reported, and its possible origin was linked to the presence of TSS coexisting with a bulk semimetallic state.[23] Further, the first angle-resolved photoemission spectroscopy (ARPES) confirmed the presence of mixed trivial and TSS in $TaAs_2$; however, the surface states were poorly visible, and the authors attributed it to the non-perfect surface cleavage.[24]

It is noteworthy that TSS nurture the potential for high-speed dissipationless electronics, spintronics, memory devices, and robust quantum bits (qubits).[1, 2] Nevertheless, identifying and extracting transport of quasiparticles associated with these TSS requires meticulous efforts to suppress overwhelming bulk electronic contribution.[2, 25] Low-dimensional systems, e.g., nanowires (NWs), nanoribbons (NRs), thin films, and vdW materials, with pronounced surface-to-bulk ratio, provide an essential platform to examine potential surface states and their contribution to transport.[1, 2] Among different low-dimensional systems, the NW one-dimensional (1D) geometry is particularly interesting owing to its ability to manifest two important phenomena: (1) Aharonov-Bohm (AB) oscillations: oscillations in the electrical resistivity in the presence of a longitudinal magnetic field, which has been observed in NWs (NRs) of TCIs (TIs), attributed to topologically protected metallic surface states,[26] and (2) Majorana zero modes (MZM) emerging at interfaces between different topological states along the wires, or as end states, in the proximity of a superconductor.[27] MZMs have been proposed as an attractive building block for realizing robust quantum computers.[27, 28, 29] Despite the anticipated rich electronic band topology and transport features in $TaAs_2$ and other analogs, an understanding of their surface states and related transport features is absent due to the unavailability of nanostructures from these materials. Thus, it is highly desirable to produce and study low-dimensional structures of this family of materials.

Besides the lack of nanostructures made of specific topological materials, like $TaAs_2$ in this case, a more general problem of topological nanomaterials is that many of them, especially pnictides and chalcogenides, are prone to oxidation, which chemically degrades their surface, and makes surface transport phenomena poorly visible.[2] It is thus important to devise a general methodology for producing nanostructures of topological materials wherein the surface is chemically protected.

In this article, we report the synthesis, structure, and magnetotransport properties of topological semimetal $TaAs_2$ NWs. These NWs are grown by a chemical vapor-assisted mechanism that *in situ* encapsulates them in a thin shell of $SiO_2$, which conveniently acts as a protective layer and dielectric for electrical gating. The $SiO_2$ shell can be locally etched to make ohmic contacts while the rest of the NW remains encapsulated. The morphology and single-crystalline structure of the NWs are examined using scanning electron microscopy (SEM), atomic resolution scanning transmission electron microscopy (STEM) with energy dispersive spectra (EDS), atomic-resolution chemical mapping, and Raman spectroscopy. NWs with varying $TaAs_2$ core diameters are integrated into four-probe devices to study their rich magnetotransport properties. The NWs exhibit a metallic conductivity that turns into an insulating state in an external magnetic field, resulting in a tunable MI transition. Notably, in the NWs, field-induced MI transition occurs at nearly four times higher temperatures than their bulk counterpart, reaching close to room temperature. Moreover, a low-temperature universal resistivity plateau known in the bulk phase is replaced by a previously unobserved non-trivial metallic conduction mode. Another effect of the magnetic field is a highly anisotropic GMR, and its tunability with $TaAs_2$ diameter. Lastly, the MR sign-reversal and quantum oscillations in a field perfectly aligned with the NW-axis are discussed. A theoretical model is proposed to qualitatively explain how these phenomena in the NWs can arise from the different topological phases and the quasi-1D geometry.

**Synthesis and structural analysis**

We synthesized $TaAs_2$ NWs employing a chemical vapor reaction of Ta and As precursors in a confined atmosphere allowing the postgrowth of a thin $SiO_2$ shell. Low-magnification SEM images show a large-area growth of NWs on different plane-sapphire (C, R, and AM-$\alpha$-$Al_2O_3$) substrates (**Fig. 1a**, **S1, S2)**. The lengths of NWs range from a few tens to a few hundred microns, and it can be noticed that some wide nanobelt (NBs) structures also form alongside thin NWs. High-resolution SEM and TEM reveal that the $TaAs_2$ NWs possess a core-shell structure wherein the single-crystalline $TaAs_2$ core is uniformly encapsulated by an amorphous $SiO_2$ shell (**Figs. 1b,c**). EDS from the core confirms the Ta:As ratio close to 1:2 (**Fig. 1c Inset**). A survey of SEM and TEM on several samples showed that $TaAs_2$ core diameters (*d*) range between 14 - 450 nm, and the $SiO_2$ shell around them is 60 - 80 nm thick. .

To study the crystal structure and orientation, atomic-resolution STEM was performed on NW cross-sections prepared by a focused-ion-beam (FIB). **Fig. 1d (**upper panel) presents the low-resolution high angle annular dark field (HAADF) STEM image of a NW cross-section, showing an irregular hexagonal $TaAs_2$ core with a {010} plane cross-section sided by parallel pairs of {001}, {201}, and $\{20\bar{1}\}$ facets, surrounded by an amorphous $SiO_2$ shell. The lower panel in **Fig. 1d** shows the EDS-elemental mapping from the core with a homogenous presence of Ta and As. Atomic resolution HAADF STEM image of the core (**Fig. 1e**) and corresponding fast Fourier transformation (FFT) pattern (**Inset Fig. 1e**) reveals the highly ordered single-crystalline nature of the NW, with (001), (100), (201), (101) lattice planes along a zone axis of [010]. The identified planes and corresponding lattice spacing agree with the reported monoclinic crystal structure of $TaAs_2$ (C2/m).[21] Note that the NW growth direction <010> is the easy crystallization direction of monoclinic $TaAs_2$.

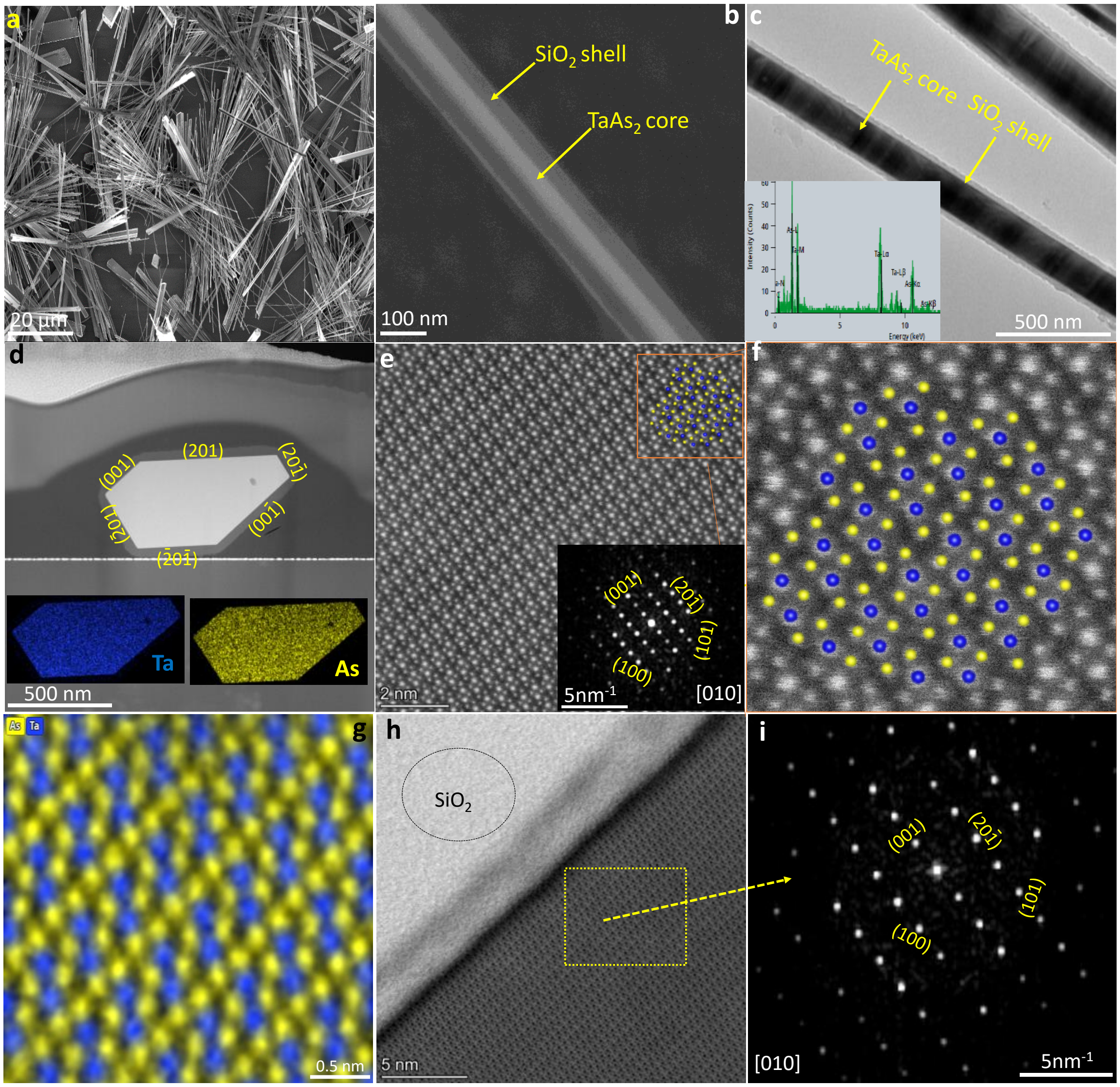

**Fig. 1 Synthesis and structural analysis of $TaAs_2$ NWs. a** Low-magnification SEM image showing large area growth of NWs. **b** High-magnification SEM image showing a single NW with core-shell $TaAs_2$@$SiO_2$ structure. **c** Top-view TEM image of a NWs showing core-shell structure; inset shows EDS spectrum from core with Ta and As peaks. **d** Top panel: HAADF-STEM image showing a cross-section of NW with $TaAs_2$ core, and exposed facets encapsulated in an amorphous $SiO_2$ shell. Bottom panels: EDS elemental mapping from the core region. **e** Atomic-resolution HAADF-STEM image from the NW core showing highly ordered $TaAs_2$ lattice; inset shows corresponding FFT pattern and lattice planes of single-crystalline core. **f** Atomic-resolution fit of NW lattice with reported structure. **g** Atomic-resolution EDS mapping of core showing lattice arrangements of Ta (blue) and As (yellow). **h** High-resolution bright-field (BF) STEM image of interface showing highly crystalline $TaAs_2$ surface with amorphous $SiO_2$ shell. **i** FFT pattern from interface region showing lattice planes of single-crystalline $TaAs_2$.

**Fig. 1f** shows the fitting between the observed and simulated crystal structure.[30] Atomic-resolution EDS mapping of $TaAs_2$ core (**Fig. 1g**) display lattice arrangements of Ta (blue) and As (yellow) atoms wherein each Ta atom surrounded with six As atoms can be seen. It can be noticed that As atoms adopt two different lattice arrangements directed by their different chemical states ($As^{3-}$ and $As_2^{4-}$). To the best of our knowledge, this is not only the first report of $TaAs_2$ NWs but also the first atomic resolution TEM of a $TaAs_2$ crystal, which shows the lattice arrangements of Ta and As atoms in the crystal (**Fig. S3**).

Next, we examined the interface between the single-crystalline $TaAs_2$ core and the amorphous $SiO_2$ shell. Atomic resolution STEM image shows a clean interface (**Fig. 1h**) with the FFT pattern under a zone-axis [010] manifesting (001), (200), (201), (101) planes corresponding to single-crystalline $TaAs_2$ surface with no signs of alloying or substitution or any impurity (**Fig. 1i**). In supporting information (**Figs. S4, S5**) we show cross-sectional STEM and crystal structure analysis from a different NW. The single-NW-based Raman spectrum from several samples shows characteristic peaks that agree with the reported $TaAs_2$ spectrum (**Fig. S6**).[30]

**Electrical and magnetotransport properties in a transversal magnetic field**

For studying the magnetotransport properties, individual NWs, as shown in **Fig. 1,** were integrated into four-terminal devices (**Fig. S7**). Contacts were made by local etching of the $SiO_2$ shell immediately followed by metal deposition (See Methods). The electrical conductivity of NWs with $TaAs_2$ core diameter (*d*) ranging from 30-450 nm exhibited Ohmic contacts, and ampacity (current carrying capacity) as high as 3 to 4 mA under ambient conditions (**Fig. S8**). **Fig. 2a** shows a typical room-temperature *I-V* characteristic curve from a NW (**Fig. 2a lower-inset SEM image**) with *d* ~300 nm. The electrical behavior of the NWs exhibits a metallic nature, with their resistivity starting to saturate near 25 K (**Fig. 2a upper inset**). Notably, all NWs with *d* range ~ 14-450 nm exhibit a metallic conductivity with their low temperature (2 K) resistivity as low as 1.7-35 μΩ·cm.

Application of a magnetic field perpendicular to the NW axis <010> and the current direction, increases the resistivity with a pronounced effect in low-temperature regions, where a transition from metallic-to-insulating (MI) state occurs (**Fig. 2b**). To better analyze the MI transition, the first derivative of the resistivity $\delta\rho/\delta T$ with respect to temperature under varying field strengths is shown in **Fig. 2c**. The $\delta\rho/\delta T$ changes sign from positive to negative due to the MI transition, and the corresponding transition temperature ($T_m$) shows a linear increase with the field strength (**Fig. 2d**). It is noteworthy that the MI transition in NWs occurs at much higher temperatures ($T_m$) than in bulk $TaAs_2$.[20, 21]

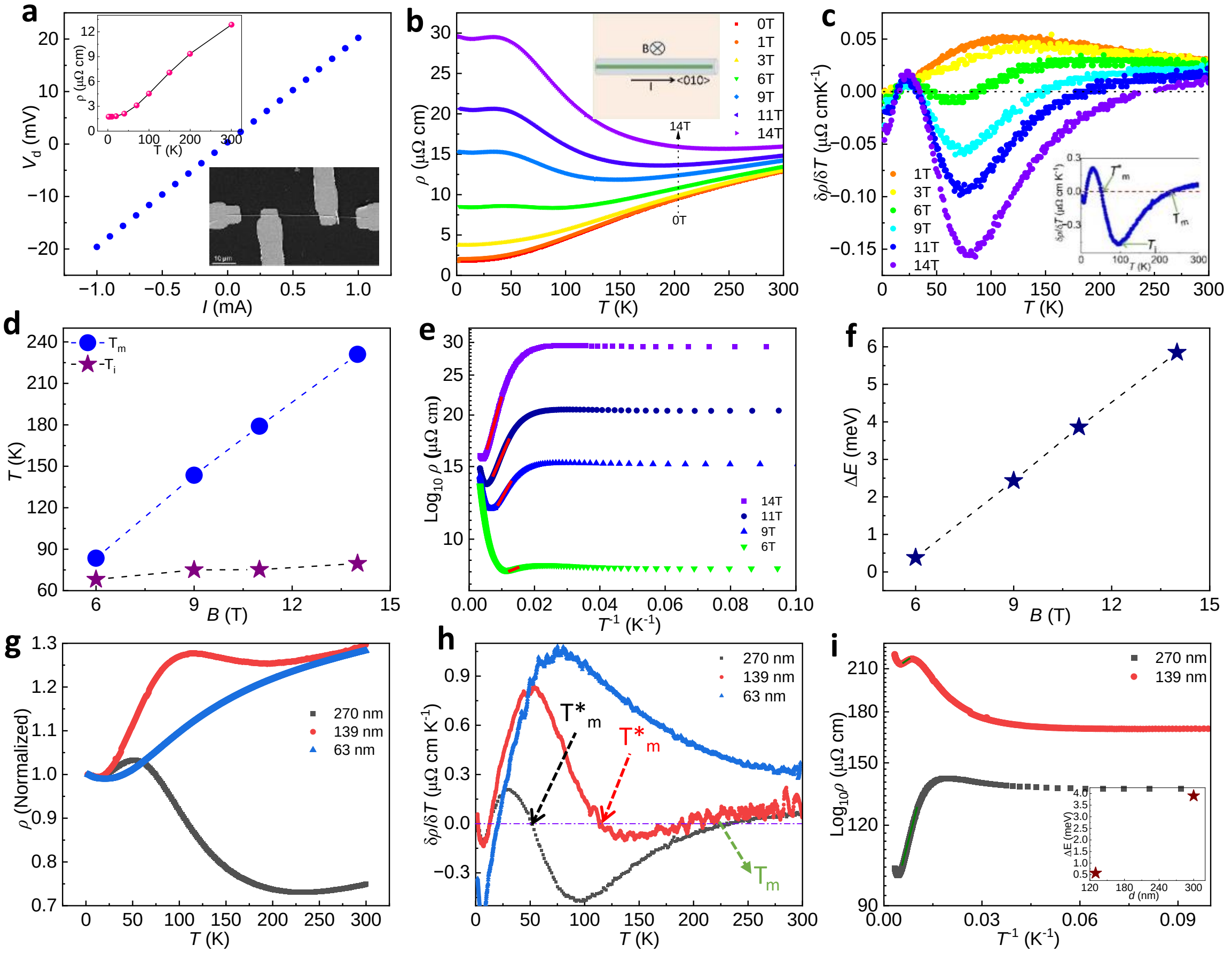

**Fig.2 Transport properties of $TaAs_2$ NWs in a perpendicular magnetic field configuration. a** A four-probe current-voltage (I-V) curve from a $TaAs_2$ NW *d* ~ 300 nm; lower inset show SEM image of the device; upper inset show resistivity vs. temperature (RT) in the absence of magnetic field. **b** RT curves of the same device under increasing field intensities (dotted black arrow represents increasing 0-14 T); Inset show schematic representation of top-view NW with applied electric and magnetic field directions. **c** Corresponding first-derivative of resistivity $\delta\rho/\delta T$ as a function of temperature; Inset show metal-to-insulator (MI) and insulator-to-metal (IM) transition as a function of temperature. **d** MI transition temperature ($T_m$) and inflection point temperature ($T_i$) obtained from **c** as a function of field intensity. **e** Arrhenius $\log_{10}\rho$ vs $T^{-1}$ plots of the insulating phase under increasing field strengths for the same device. **f** Obtained insulating gaps from **e** as a function of magnetic field strength. **g** *RT* curves for NWs with varying $TaAs_2$ core *d* ~ 270, 139, and 63 nm at 14 T field. **h** Corresponding $\delta\rho/\delta T$ vs. temperature plots showing MI ($T_m$) and IM ($T^*_m$) transitions with respect to NWs diameter. **i** Arrhenius $\log_{10}\rho$ *vs.* $T^{-1}$ plot for NWs (*d* ~ 270 and 139 nm); inset shows recorded insulating gap in these NWs at 14 T.

Notably, at 14 T, $T_m$ reaches 236 K for a *d* ~ 300 nm NW device, which is almost 3.5 times higher than the transition temperature (~60-70 K) reported in bulk $TaAs_2$.[22] It is technologically appealing that metallic $TaAs_2$ NW devices can be tuned into an insulating state near room temperature.

The insulating phase under a magnetic field originates from the formation of cyclotron orbits of carriers along and across the NWs, thus producing quantized energy levels, i.e., Landau levels.[31] Insulating gaps under varying fields at $T \leq T_m$ can be obtained using the Arrhenius equation $ln\,\rho\,(T) = \ln K + \Delta/K_B T$ (**Figs. 2e, S9**). The linear region fit of a ln $\rho$ vs. $T^{-1}$ provides slopes equivalent to $\Delta/k_B$, where $\Delta = \hbar\omega_c - \Gamma = \hbar\ (eB/m^*) - \Gamma$ is the insulating gap linked to the Landau energy levels, $\omega_c = eB/m^*$ is the cyclotron frequency, $e$ is the elementary charge, $m^*$ is the effective mass, $B$ is field strength, $\Gamma$ is a localization parameter associated to the boundary of NWs, $k_B$ is Boltzmann constant, and $K$ is a constant value. The insulating gaps ($\Delta$ in meV) show a linear dependence and increase with the field strength as expected from the above equation (**Fig. 2f**). Note that localization effects are stronger in NWs than their bulk-counterparts hence, $\Gamma$ is expected to be larger in NWs.[32] Therefore, an insulating phase in NWs can only occur if $\hbar(eB/m^*) > \Gamma$, and the resulting gap will always be smaller than bulk counterparts. This explains the smaller gap (5.8 meV) observed in NWs ($d$ ~ 300 nm) with respect to bulk $TaAs_2$ (15.3 meV).[22]

Notice that with a further decrease in temperature, $\delta\rho/\delta T$ reached its minimum, followed by a swift change at the inflection point ($T_i$) (**Fig. 2c, inset**). In bulk $TX_2$ and other semimetals (LaSb, $MoAs_2$, $WTe_2$), $T_i$ was used as a marker of the origin of the resistivity plateau.[33, 34, 35] In $TaAs_2$ NWs, we observe replacement of the resistivity plateau by a metallic conduction mode manifested by a second change in $\delta\rho/\delta T$ from negative to positive, thus giving rise to a non-trivial insulator-to-metal (IM) transition at $T^*_m$ (**Fig. 2c and inset)**. Furthermore, magnetotransport features at 14 T in NWs with varying $TaAs_2$ core ($d$ ~ 270, 139, and 63 nm) demonstrate the field-induced electrical transitions as a function of temperature (**Fig. 2g**). It can be noticed that features change markedly in different NW core $d$, with 270 nm exhibiting MI transition at $T_m$ ~ 236 K. The corresponding $\delta\rho/\delta T$ manifests magnetic field-induced MI and IM transitions (**Fig. 2h**). Notably, the replacement of the resistivity plateau by the metallic conduction mode at $T \sim T_i$ is enhanced by the higher surface-to-bulk ratio in NWs, and the corresponding non-trivial IM transition temperatures ($T^*_m$) increase with the core $d$ reduction. A shift of $T^*_m$ from 51 to 115 K due to a reduced $d$ (from 270 to 139 nm) is noteworthy (**Fig. 2h**). In $d$~ 63 nm NW, we did not observe any electrical transition, and it maintained a metallic nature even at 14 T.

In TSMs, a universal resistivity plateau has recently been reported under broken time-reversal-symmetry (TRS) in an external magnetic field.[33, 34, 35] The resistivity plateau in LaSb was indicated to originate from TSS.[33] The authors provided an analogy with a similar plateau in a TI $SmB_6$ under a zero field where TRS was preserved. In $MoAs_2$, the appearance of a similar field-induced resistivity plateau saturating the bulk insulating resistivity at low temperatures was also attributed to TSS transport contribution.[34] A similar plateau has also been observed in bulk $TX_2$ compounds; in $TaSb_2$, it was attributed to the presence of metallic TSS based on the observation of a non-trivial Berry phase.[36]

We show in quasi-1D $TaAs_2$ NWs that pronounced surface states lead to the replacement of the resistivity plateau by a dominating metallic surface conduction mode, leading instead to the appearance of a non-trivial IM transition. In light of the rich coexisting electronic band topologies and associated TSS, e.g., WTI,[6, 7] $C_2$-symmetry-protected TCI,[16] and Zeeman field-induced Weyl points,[7] it is likely to observe a strong TSS expression, particularly in NWs which could enable the strong manifestation of these TSS due to high surface vs bulk transport ratio. Since our NW terminates at the {010} surface, the observed TSS transport cannot be associated with the TCI phase, which is predicted to host type-II Dirac fermions on that surface.[16] However, the WTI phase of $TaAs_2$ is predicted to have a metallic surface state on the {001}, {201}, and $\{20\bar{1}\}$ surfaces (**Fig. 1e**) that can potentially contribute to the observed surface conduction.[6] It is noteworthy that despite being called "weak" TI, the surface states of WTIs actually possess predicted robust features, which brings about richness in the transport.[8] Also, the surface Fermi arcs associated with Zeeman field-induced bulk Weyl cones cannot be ignored as a potential source of surface conduction[7]. It is of great technological significance that quasi-1D $TaAs_2$ NWs maintain TSS transport features at notably higher temperatures (**Fig. 2h**) than known topological semimetals.

Returning to the discussion on the insulating state, the gaps linked to the MI transition in thinner NWs decrease (**Fig. 2i, inset**) due to an increase in localization parameter ($\Gamma$), which would enforce a stronger magnetic field to obtain the condition $\hbar(eB/m^*) > \Gamma$ as discussed above.[32] Magnetotransport studies performed on a number of NWs manifested robust transport features (**Figs. S10, S11**). An additional insulating state was observed in all NWs at temperatures $T \leq 12$ K, likely due to the Kondo effect at large fields[37] or oscillating behavior of resistivity due to strong Shubnikov-de Haas (SdH) effect, although further analysis would be needed to demonstrate this.

## Giant magnetoresistance (GMR) in a transversal magnetic field

A second consequence of TRS breaking is the occurrence of a field-dependent GMR. In **Fig. 3a,** we present the MR (% $MR = (\rho_B-\rho_0)/\rho_0\times100$) measured at 2 K in two independent devices NW (NB) with $TaAs_2$ core diameter (thickness) ~ 300 nm. A non-saturating GMR of the order $10^3$ % was observed in both devices. The log $MR$ vs log $B$ plot yielded a linear dependence with a slope of 1.8±0.05, indicating a close to quadratic dependence (**Inset Fig. 3a**). Note that the quadratic dependence of MR in $TaAs_2$ NWs is consistent with the similar dependence observed in the bulk $TX_2$, associated to the nearly perfect electrons and holes compensation.[21]

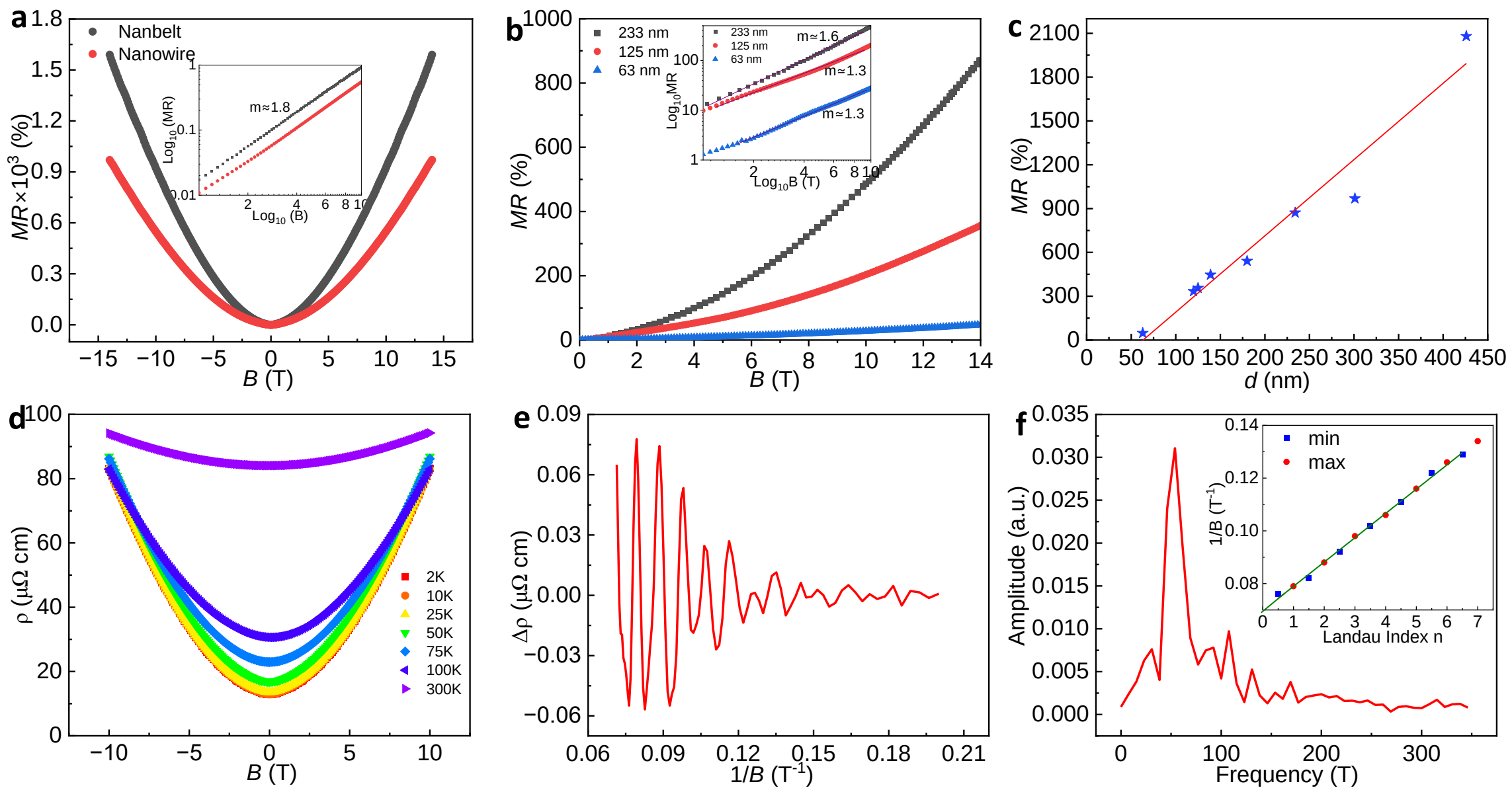

**Fig.3 Giant Magnetoresistance (GMR) and Quantum Oscillations in a perpendicular field configuration. a** Field dependent magnetoresistance (MR) in a $TaAs_2$ nanowire (nanobelt) of 300 nm thickness (diameter) at 2K; inset shows log *MR* vs. log *B* plot with a quadratic field dependence. **b** MR in NWs with varying $TaAs_2$ core, d ~ 233, 125 and 63 nm; Inset shows log *MR* vs log *B* plots and obtained slops (m). **c** MR as a function of NW core diameter *d* (red solid line represent linear fit). **d** Temperature dependence of MR in a NW (*d* ~ 270 nm). **e** Shubnikov-de Haas (SdH) oscillations from the same device at 2 K. **f** FFT spectrum of SdH-oscillations; inset shows Landau fan diagram with blue (red) circles representing $\Delta\rho$ min. (max.) respectively.

In compensated semimetals, MR is derived from the simplification of semiclassical two-band theory and maintains quadratic dependence even under strong fields as MR= $\mu_e\mu_h B^2$ wherein $\mu_e$ and $\mu_h$ are the mobilities of electrons and holes, respectively, and *B* is the field strength.[20, 35] To further understand the behavior of the GMR, we measured MR in devices with varying $TaAs_2$ core *d* (**Fig. 3b**). The MR of NWs with *d*~ 233, 125, and 63 nm decreases unequivocally with the decrease in the core *d*. A log *MR* vs log *B* yields a straight line of slope 1.6 ± 0.03 for *d*~ 233 nm and 1.4 -1.3 for $TaAs_2$ *d* ~ 125 and 63 nm, respectively (**Inset Fig. 3b**)**,** whereas in a *d*~ 300 nm $TaAs_2$ core, the slope was 1.8 ± 0.05 (**Inset Fig. 3a**). Such a slope decrease suggests the weakening of semiclassical quadratic field dependence of MR in thinner NWs. **Fig. 3c** summarizes the dependence of the maximal MR on $TaAs_2$ *d* (red solid line is the linear fitting); the MR increases almost linearly with the increase in *d* and spanning from 47 % (*d* ~ 63 nm) to 2100 % (*d*~ 420 nm). The decrease of the MR in thinner NWs can be explained by the correlation between the cyclotron orbit diameter with respect to the NW *d* in an external magnetic field. It has been shown in compensated semimetals, e.g., Bi and Sb NWs, that the MR exhibits a maximum at the field where the cyclotron radius becomes roughly equal to the NW's radius, and below this field, surface scattering dominates.[38, 39] Therefore, thinner NWs would need a stronger field translated into a smaller cyclotron orbit to produce MR of the same magnitude. Notably, GMR observed in $TaAs_2$ NWs of *d* in the range of 250-400 nm are among the highest in the reported trivial and topological semimetals/insulator NWs.[25, 39, 40, 41] Note that the MR in $TaAs_2$ NWs also shows a robust behavior against temperature, and decays very slowly up to 100 K (**Fig. 3d, S12a**), unlike in bulk phase, where it drops orders of magnitude under the same conditions.[22] A MR of 13-15 % was recorded in NWs of *d* ~ 250-350 nm even at room temperature (**Figs. 3d, S12a**).

The GMR in $TaAs_2$ NWs can further be tuned by electrical gating, wherein the naturally grown $SiO_2$-shell works as an efficient gate dielectric. Thus, the *in situ* grown shell around the $TaAs_2$ core has a double benefit: it acts as a removable protective layer for electrical contact and a dielectric layer for gating. Gating experiments are presented in **Fig. S13** and **SI** discussion.

The quantum oscillatory part of MR is characterized by subtracting the semiclassical background (**Fig. 3e, S12b)**. Oscillatory features at 2 K are characteristic of the Shubnikov-de Hass (SdH) effect.[22] The rapid periodic beating of the resistivity indicates Landau quantization of complex Fermi surfaces with different high-mobility carriers. Unlike two major oscillation frequencies observed in bulk $TaAs_2$,[22] the FFT of oscillations in NWs manifest one major frequency peak at 53 T (**Fig. 3f**). The corresponding Fermi surface area ($A_F$) normal to the applied field was deduced employing the Onsager relation, $F = (\Phi/2\Pi^2)A_F$, wherein $F$, $\Phi = (h/2e)$, and $A_F$ are the FFT peak position, flux quantum, and Fermi surface cross-sectional area, respectively. Thus, the calculated $A_F = 0.50 \times 10^{-2}$ (Å$^{-2}$) is in agreement with ARPES results on bulk $TaAs_2$.[24] Furthermore, the Lifshitz-Onsager quantization equation $A_F \frac{\hbar}{eB} = 2\Pi\,(n + ½ + \beta + \delta)$ can be used to identify trivial (0) or non-trivial ($\Pi$) Berry phase, where $n$ is Landau-level index, $2\Pi\beta$ is the Berry phase and $2\Pi\delta$ is a parameter linked to the Fermi surface curvature with its value $\delta = \pm 1/8$ for a 3D Fermi surface.[3] The inset in **Fig. 3f** shows the Landau fan diagram, with integer ($n$) and half-integer ($n$ +1/2) indices representing $\Delta\rho$ maximum (peak) and minimum (valley), respectively. The linear dependence of Landau index $n$ on $1/B$ gives $x$-intercept = 1/2 + 1/8+ $\beta$ = 0.05 ± 0.01 and thus $\beta = 0.57 \pm 0.01$ and a Berry phase of ($2\Pi\beta$) = $\Pi \pm 0.1$, a characteristic of Dirac fermions linked to the topologically non-trivial Fermi surface.[3]

**Anisotropic and Negative Magnetoresistance**

We next examined the magnetotransport in a $TaAs_2$ NW (*d* ~ 270 nm) under different angular orientations ($\theta$) between the NW axis and field direction (**Inset Fig. 4a**). The MR shows a strongly anisotropic behavior, where the GMR decreases with decreasing the angle between the NW axis and field direction (**Fig. 4a, S14a**). A notable downward cusp of the resistivity near zero fields becomes shallow and broadens as the angle approaches $\theta \sim$ 5-6°, followed by a positive MR turning into a negative at higher fields (**Figs. S14b, S15a**). This angle dependence of the observed cusp (**Fig. S15b**) indicates a sign of 2D weak antilocalization (WAL), expected from the topological surface states.[42] In a relatively thinner NW ($d$ ~128 nm) under a perfectly longitudinal field configuration ($\theta$ = 0°), an unambiguous longitudinal negative magnetoresistance (LNMR) was observed in the entire measured field range (**Fig. 4b**). The field dependence of LNMR fits equation $\rho = \rho_0 - 0.020(B)^2$ wherein $\rho$, $\rho_0$, and $B$ represent field-dependent resistivity, zero-field resistivity, and field strength respectively (**Fig. S16**). LNMR of NWs aligns with a similar effect reported in bulk $TaAs_2$.[23] The LNMR in $TaAs_2$ and other $TX_2$ analogs potentially originate from the chiral charge pumping (Weyl fermions transport) between the Weyl nodes of opposite chirality induced by Zeeman splitting of bulk bands (**Fig. 4j**).[7, 36, 43] The presence of topologically non-trivial Fermi surface with $\Pi$ Berry phase (indicative of Dirac/Weyl fermions) in $TaAs_2$ NW is in support of this argument.[44]

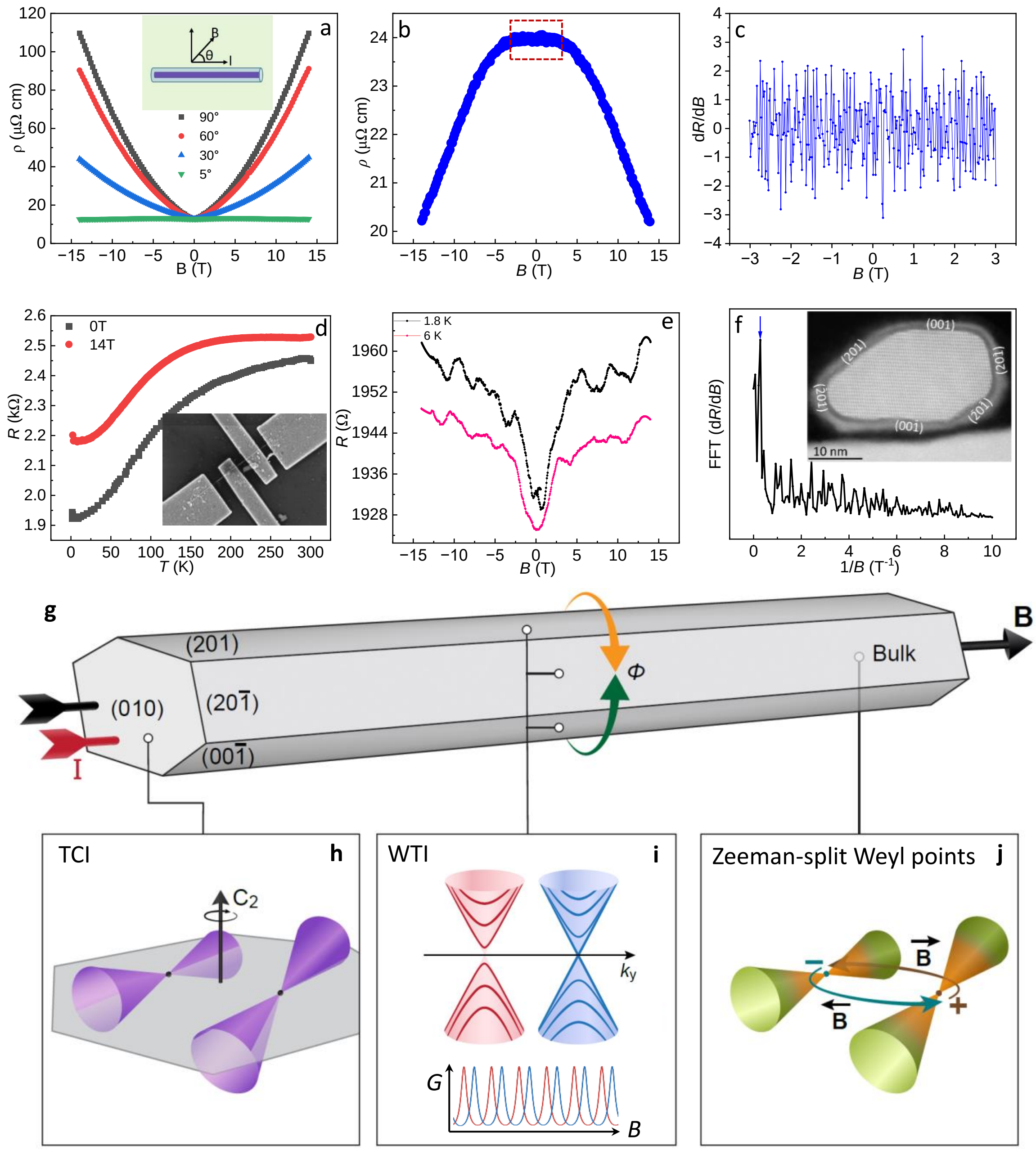

**Fig.4 Anisotropic Magnetoresistance, Aharanov-Bohm (AB) oscillations and intertwined topological states in $TaAs_2$ NWs. a** Resistivity at 2 K as a function of magnetic field under different orientations of a NW (*d* ~ 270 nm) with respect to the field direction; inset shows top-view schematic presentation of NW-axis rotation with respect to field. **b** Negative MR in a NW (*d* ~ 128 nm) under perfect longitudinal field configuration ($\theta \sim 0°$) at 2 K. **c** MR oscillations/fluctuations for the same NW device in a field (dotted square in **b)** range (±3 T). **d** Resistance as a function of temperature for the *d* ~ 32 nm NW (inset: SEM image of device) in the absence and presence of 14 T filed at $\theta = 90°$. **e** Longitudinal MR and AB oscillations at 2 and 6 K in the same NW (*d* ~ 32 nm) with a field aligned in the NW-axis ($\theta = 0°$). **f** FFT spectrum of **e** obtained after non-oscillating MR background subtraction; inset shows TEM image of the same NW cross-section with facets harboring weak topological insulator (WTI) states. **g** Schematic representation of a $TaAs_2$ NW with topologically non-trivial facets enabling quantum coherent surface transport in a longitudinal magnetic (**B**) and electric (**I**) field configuration ($\theta \sim 0°$). **h** Topological-crystalline-insulator (TCI) state with a pair of $C_2$-symmetry protected overtitled type-II Dirac cones on the (010) surface. **i** A pair of type-I Dirac cones

(light red and blue) associated with the WTI state on the {001}, {201}, and $\{20\bar{1}\}$ surfaces of the NW wall, quantized into discrete subbands (bright red and blue lines) by the quasi-1D compact geometry of the NW; the inset shows a schematic representation of AB oscillation doublet peaks originating from the superposition of two phase-shifted interference patterns arising from the two Dirac cones. **j** Zeeman-split pair of bulk Weyl cones with opposite chirality ("+" and "-"), with opposite directions of chiral charge pumping (green and brown arrows) controlled by the orientation of the electric and magnetic field, resulting in the observed longitudinal negative magnetoresistance (LNMR).

## Magnetoresistance oscillations/fluctuations in a longitudinal field

To probe the coherent electron transport linked to surface Dirac fermions[26], oscillations were measured in the same NW (*d* ~128 nm). Multiple peaks with a period close to 0.1T were recorded (**Fig. 4c**); however, the FFT spectrum (**Fig. S17)** obtained by subtracting the LNMR background (d*R*/d*B*) did not show a clear periodicity. This can be a sign of universal conductance fluctuations, which occur due to the bulk semimetallic states coexisting with the surface states,[24] resulting in a lack of a perfect electron coherent path at the measured temperature.

To further enhance the contribution of the surface and decrease that of bulk electronic states, a device was fabricated on a much thinner NW (*d* ~ 32 nm), with metallic conductivity and no apparent electrical transitions even in a transversal magnetic field of 14 T (**Fig. 4d**), indicating a situation of dominant surface transport due to much smaller NW diameter with respect to cyclotron orbit.[38] The absence of an LNMR further confirmed that transport here is dominated by surface states.[38, 39] Note that this thinner $TaAs_2$ NW not only shows sign-reversal of MR but also manifests much stronger longitudinal MR oscillations, indicating a significant expression of TSS transport (**Fig. 4e**). The period of oscillation ($\Delta B$) remains roughly constant in the entire field range with the oscillation amplitude decaying at higher temperature. The FFT d*R*/d*B* gives a major oscillation peak at ~ 0.25 $T^{-1}$, associated with the $\Delta B$ ~ 4.1 T (**Fig. 4f**) with no apparent second harmonic signal. The AB oscillation period that one would expect for a TI NW of this diameter should be $\Delta B = \Phi_0/A$, where $\Phi_0 = h/e$ is the flux-quantum, *h* is Plank's constant, *e* is elementary charge, and *A* is the NW cross-section area.[26] Considering that the NW cross-section measured by TEM is $A = 0.5\times10^{-15}$ $m^2$ (**Fig. 4f inset**), the oscillation period thus expected would be $\Delta B$ = 8.3 T. This value is about 2 times larger than the experimental value $\Delta B$ ~ 4.1 T, corresponding to a $\Delta B$ period of $h/2e$ (divided by *A*) rather than $h/e$.

The doubling of the oscillation frequency can *a priori* be attributed to two effects: (i) Altshuler-Aronov-Spivak (AAS) oscillations, known to occur in diffusive metallic cylinders and have a period of $h/2e$,[26] and (ii) a double AB oscillation pattern stemming from the fact that our wire is a WTI[6, 7, 8] and not a strong TI, and hence it has two surface Dirac cones instead of one, each contributing its own AB oscillation pattern. We argue that ASS oscillations are a less likely mechanism in our wire, for several reasons. First, the measured conductance is $5.1\times10^{-4}$ S (**Figs. 4e, S19**), which is ~ 13 times higher than the conductance quantum $G_0= e^2/h$. Such large conductance, especially for a thin NW whose magnetoresistance indicates that transport is dominated by surface states, is indicative of ballistic rather than diffusive transport. Second, the $TaAs_2$ NWs studied in this work possess a highly crystalline chemically protected surface with no traces of disorder (**Fig 4f** inset, and structural characterization), and are p-doped (**Fig. S13** and **SI** discussion), resulting in conductance much higher than $e^2/h$ and a maximum of magnetoconductance at zero flux (**Fig. S19**), compatible with AB oscillation.[45] Third, AAS oscillations are expected to become more pronounced in thicker NWs.[46] However, a thicker 128 nm diameter NW device (**Figs. 4c, S17**) did not show any oscillations with a well-defined period. Fourth, detecting a signal contributed solely by AAS oscillations is rare in the local configuration of a conventional four-probe as ours (**Fig. 4d inset**), and is usually observed only in non-local configurations.[47]

A double AB oscillation pattern (ii), on the other hand, would be consistent with the two Dirac cones of a WTI surface instead of the single Dirac cone of a TI, as schematically represented in **Fig. 4i**. WTIs are predicted to feature a pair of type-I Dirac cones on the surfaces relevant to our NW configuration[6, 8], namely parallel pairs of {201}, {001}, $\{20\bar{1}\}$ facets (**Fig. 4f inset and Figs. 4g, i**). The two Dirac cones of the WTI reside on different points of the surface Brillouin zone and can thus generate two AB oscillation patterns with the same period, but a phase shift between them (see **Figs. 4i**, **S18** and **SI** text for a theoretical model of this effect). Since each AB oscillation pattern has a period $h/e$, a superposition of the two AB oscillation patterns can give the appearance of a $h/2e$ period, as measured in our device (**Fig. 4e**), instead of $h/e$.

In quantitative terms, a conductance of ~$13G_0$ suggests that 13-14 surface subbands reside below the chemical potential. Considering that the subband $k$-spacing ($\delta k$) is the inverse of the radius, namely $(\pi/A)^{1/2}$, and the calculated Fermi velocity ($v_F$) is $3\times10^5$ m/s,[24] the energy gaps ($\delta E = \hbar v_f \delta k$) between the subbands quantized by the quasi-1D geometry are 16 meV. If the subbands contributing to this conduction stem from one or two Dirac cones, as expected from a TI or a WTI, respectively, this would be consistent with a Fermi energy difference from the Dirac cones of 200 meV or 100 meV, respectively. The latter value is very close to that reported for this energy difference in $TaAs_2$, namely 110 meV.[24] This quantitative consistency further supports ballistic conduction by a WTI surface. Systematic corroboration of this effect in NWs of $TaAs_2$ and other $TX_2$ materials with well-defined topologically non-trivial facets will pave the way to understanding and utilizing these topologically protected electrons in quantum nanotechnology.

## Conclusions

We synthesized NWs of a TSM $TaAs_2$ with intertwined topological states harboring rich electronic band topologies. These NWs possess a unique core-shell structure, with an amorphous $SiO_2$ shell encapsulating the highly crystalline $TaAs_2$ core, protecting it from oxidation or any other damage, giving remarkable ambient stability, and can be locally etched conveniently and selectively prior to electrode deposition, while leaving the rest of the NW encapsulated. The *in situ* grown shell not only protects the rich topological surface but also works as an all-around gate dielectric for tuning the Fermi energy. These high-quality NWs exhibit metallic conductivity with an impressively high ampacity of 3-4 mA under ambient conditions desirable for thermoelectrics and circuit interconnects. Introducing an external magnetic field perpendicular to the nanowire axis <010> gives rise to an MI transition near room temperature, which is four times higher than the reported transition temperature in bulk phase $TaAs_2$. Further, at $T \leq T_i$, the insulating state is replaced by a metallic surface conduction mode, resulting in a non-trivial IM transition associated with TSS. The $TaAs_2$ NWs maintain TSS transport features at remarkably high temperatures, which is technologically important for spintronic devices and qubits. Furthermore, a GMR of the order of $10^3$ increasing linearly with the NW core diameter has been achieved. Such a nonsaturating GMR in NWs could have applications in magnetic switches, memory devices, and sensors. The SdH oscillations and the Landau fan diagram demonstrate the topological Fermi surface hosting Dirac fermions. The LNMR in the $TaAs_2$ NW supports the origin of a chiral anomaly due to Zeeman field-induced Weyl points. The observation of a double AB oscillation pattern, potentially due to the occurrence of a pair of Dirac cones associated to the WTI state, opens the path to further exploration of a coherent WTI surface. We show that $TaAs_2$ NW with a thin protective shell of $SiO_2$ reported in this work serves as an excellent system to observe rich transport phenomena originating from intertwined electronic band topologies, namely a pair of type-I surface Dirac cones linked to WTI state, type-II Dirac cones due to TCI state, and Weyl points. **Figs.4g-j** summarizes the intertwined topological phases hosted by the different NW facets and their bulk, which manifest themselves in the presented results. The incorporation of superconductivity coexisting with rich TSS in $TaAs_2$ NWs can potentially provide a platform to generate Majorana fermions for topological quantum computers.[48] The protective $SiO_2$ around the $TaAs_2$ NWs makes them chemically robust and compatible with CMOS technology. Furthermore, the synthesis demonstrated here could be extended to obtain surface-clean, naturally encapsulated single-crystalline NWs of other dipnictides, like $TaP_2$, $NbP_2$, $TaSb_2$, and $NbSb_2$, $VP_2$, $VAs_2$, $VSb_2$ thus enabling the exploration of rich topological phases, exotic physical phenomena, and their potential applications in quantum nanotechnology as spintronics, magnetoelectric sensors, and quantum computing.

## Online Content

Any methods, additional references, Nature Portfolio reporting summaries, source data, extended data, supplementary information, acknowledgments, peer review information; details of author contributions and competing interests; and statements of data and code availability are available at

# Methods

## Synthesis and Charcterization of the Nanowires

$TaAs_2$ NWs were synthesized using vapor phase reaction of Ta and As precursors in an engineered condition allowing the post-growth of a thin $SiO_2$ shell.

The morphology of nanowires/nanoribbons was examined in a scanning electron microscope (Sigma 500 SEM, Zeiss). The core-shell structure of NWs/NRs was examined in a high-magnification SEM (Sigma 500 Zeiss) at an accelerating voltage of 10 KV and further confirmed using TEM (Talos F200X G2 TEM). Thin electron-transparent cross-sections of nanowire/nanoribbon were obtained by focused-ion-beam (FBI, FEI Helios 600, Dual-beam microscope). Crystal structure, atomic resolution imaging, EDS mapping, and interface imaging were done using a high-resolution transmission microscope (HR-TEM, Themis-Z). Raman spectrum of the samples was recorded using a micro-Raman spectrometer (Horiba LabRAM HR evolution). A 532 nm green laser was aligned and focused on a single NW/NR specimen under an optical resolution of X150.

## Device Fabrication and Transport Measurements

Four-terminal devices were fabricated from NWs on $\alpha$-$Al_2O_3$ or Si/$SiO_2$ (300 nm thermal oxide layer) substrates using either laser-assisted lithography or electron beam lithography. A 20 nm of $HfO_2$ layer on Si/$SiO_2$ substrates was deposited using atomic layer deposition (ALD) to protect the $SiO_2$ layer of the substrate against the buffer oxide etchant (6:1 volume ratio of 40 % $NH_4F$ in $H_2O$ to 49 % HF in $H_2O$). After the process of lithography, the $SiO_2$-shell on NW was selectively etched in a buffered oxide etchant (BOE 6:1) (30-60 seconds) to expose the $TaAs_2$-core, and immediately an electron beam evaporation of Cr/Au (10/250 nm) was done on the etched area of NW.

The resistivity magnetotransport measurements were performed in a Quantum Design 14 T PPMS Dynacool system, using the electrical transport and the resistivity modes. These options generate either a DC or an oscillating low-frequency current and measure the voltage across the device using a four-probe wiring scheme. Currents of up to 1 mA were used, with frequencies around 0.5 Hz in the electrical transport option. The angle-dependent measurements were done using Dynacool's horizontal rotator option. Back and top-gate voltage was applied using a Keithley 2450 source meter.

## Acknowledgments

A.R. acknowledges a Faculty Dean's postdoctoral fellowship. We thank Iddo Pinkas and Yarden Daniel for assistance with Raman spectroscopy, Eran Mishuk for help with ALD, and Reshef Tenne for providing synthetic facilities. This research was supported by the Israel Science Foundation (ISF) grant No. 1045/23. E.J. holds the Drake Family Professorial Chair of Nanotechnology. This research (RI and RMS) is supported by a grant from the US-Israel Binational Science Foundation (BSF, No. 2018226), Jerusalem, Israel.

## Author Contributions

E.J. and A.R. conceived the idea of the work, A.R. synthesized, characterized $TaAs_2$ NWs and fabricated devices for magnetotransport measurements. A.E. and A.R. performed the magnetotransport measurements, R.M.S., and R.I., provided the theoretical interpretation of the results, including the model and calculation for the AB-oscillations. B.B., did gate dependent MR measurements and measurements on thin NWs, S.D.E., read the manuscript and provided theoretical insights on the discussion about the insulating state behavior and Landau quantization in NWs, and suggested presenting the Landau Fan diagram. O.B. performed TEM measurements and facet identification, K.R. performed FIB to produce electron-transparent NW-cross-sections. O.B., fabricated devices on very-thin NWs using electron-beam lithography, E.J. helped in manuscript writing and suggested various experiments. A.R., wrote the manuscript, did the analysis of results and characterizations.

## Competing Interests

The authors declare no competing interests.

## Additional Information

**Supplementary Information** The online version contains supplementary material available at

**Correspondence and request for materials** should be addressed to Ernesto Joselevich and Anand Roy.

Supporting Information for

# Intertwined topological phases in $TaAs_2$ nanowires with giant magnetoresistance and quantum coherent surface transport

Anand Roy[a*], Anna Eyal[b], Roni Majlin Skiff,[c] Barun Barick[d], Samuel Díaz Escribano[d], Olga Brontvein[e], Katya Rechav[e], Ora Bitton[e], Roni Ilan[c] and Ernesto Joselevich[a*]

[a]Department of Molecular Chemistry and Materials Science, Weizmann institute of science, Rehovot 7610001, Israel; [b]Physics department, Technion, Haifa, 32000, Israel; [c]Raymond and Beverly Sackler school of Physics and Astronomy, Tel Aviv 69978, Israel [d]Department of Condensed Matter Physics, Weizmann Institute of Science, Rehovot 76100, Israel. [e]Chemical Research Support, Weizmann Institute of Science, Rehovot 76100, Israel. e-mail: ernesto.joselevich@weizmann.ac.il, anand-kumar.roy@weizmann.ac.il.

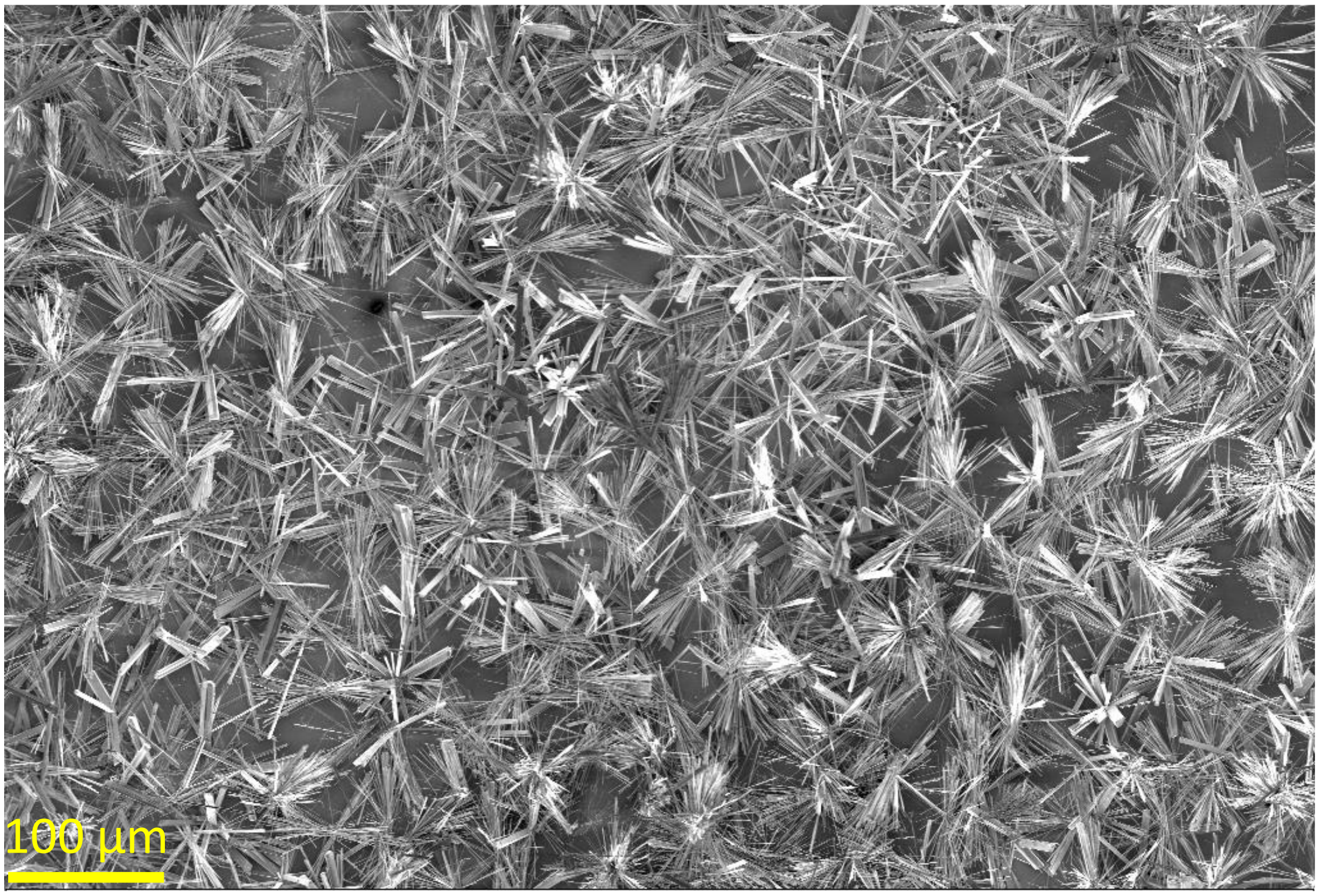

**FigS1.** Large-area SEM image showing high yield of $TaAs_2$ nanowires (NWs) with some nanobelts (NBs) on a $\alpha$-$Al_2O_3$ (sapphire) substrate.

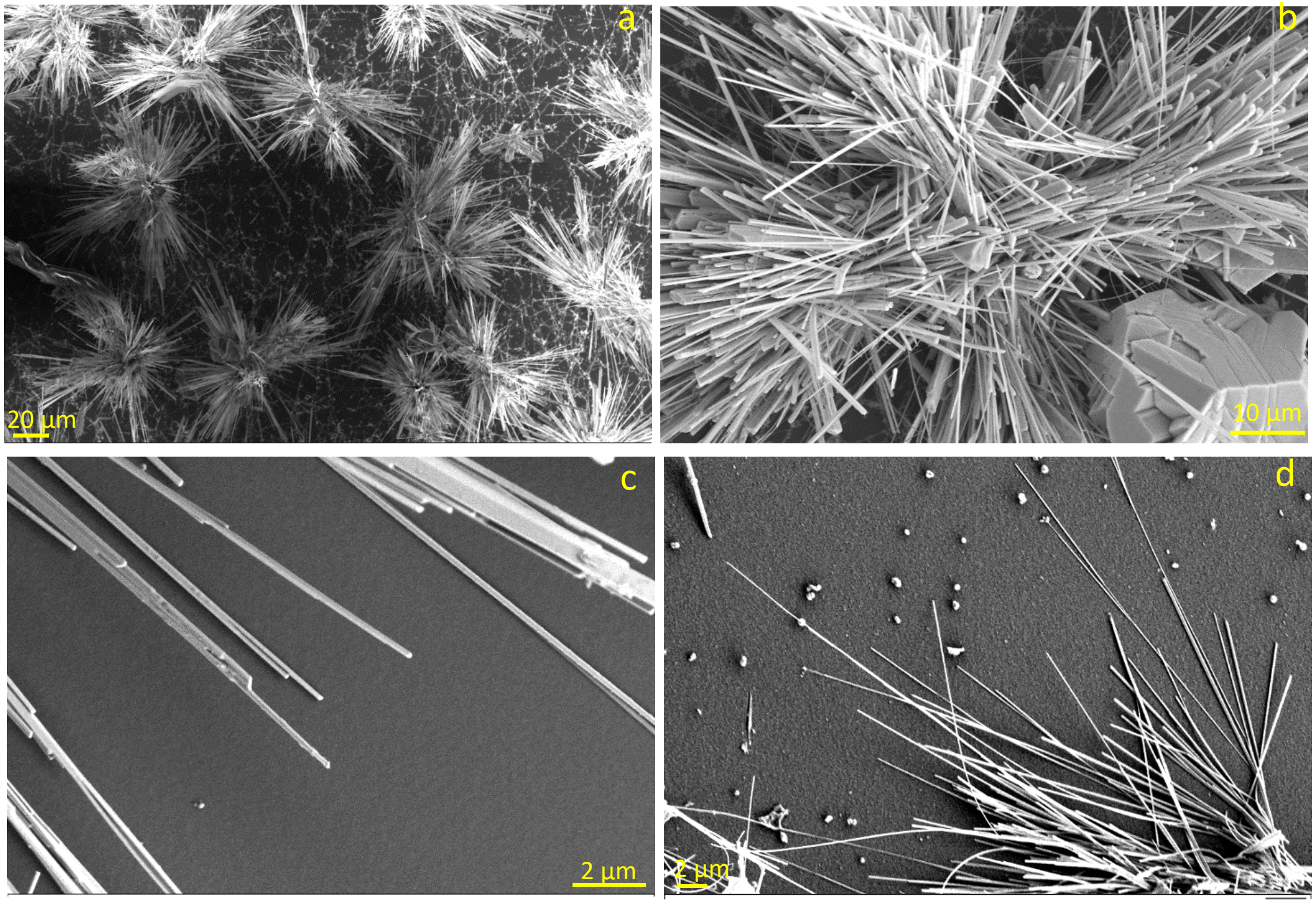

**FigS2.** SEM images of $TaAs_2$ NWs grown of different cut (R, C, and annealed-M) $\alpha$-$Al_2O_3$ substrates.

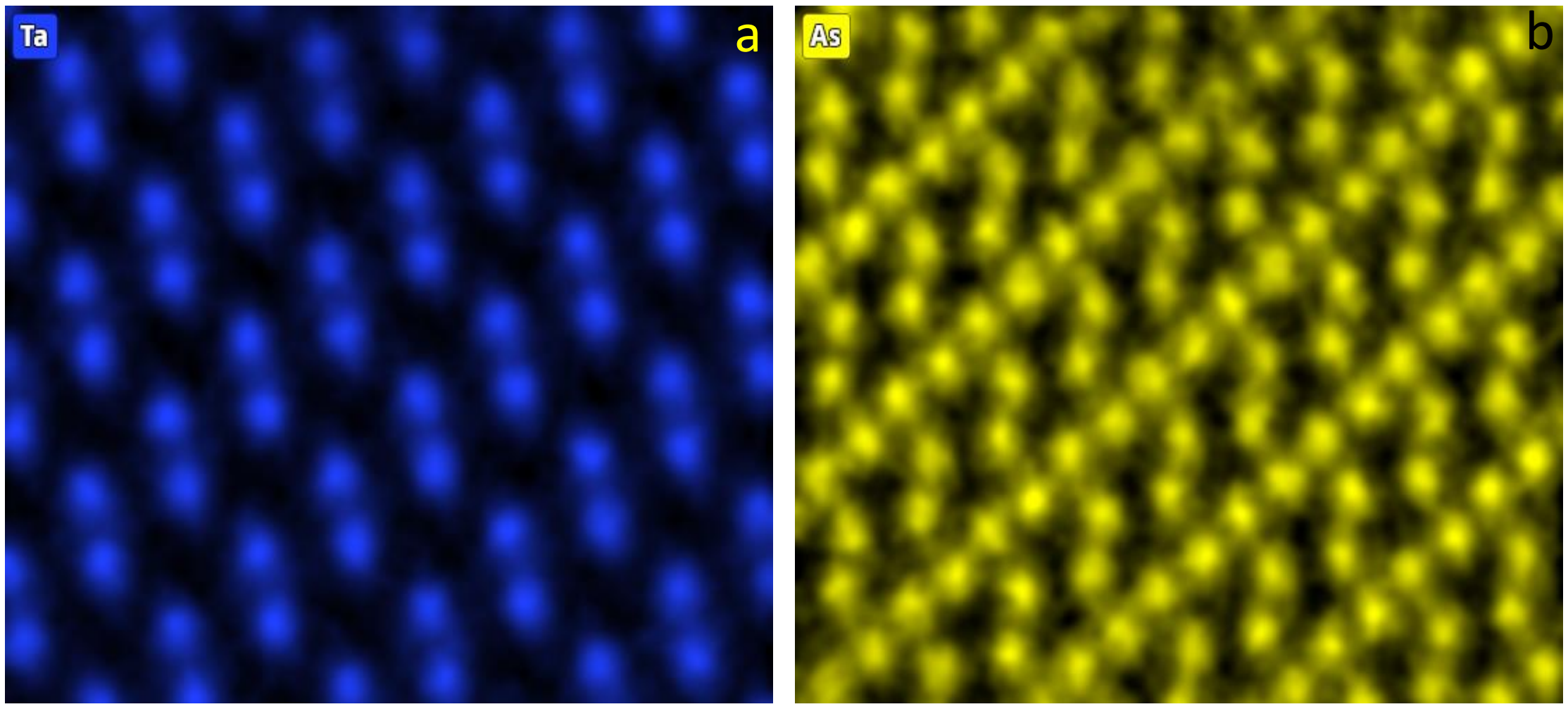

**FigS3.** Atomic resolution EDS mapping of $TaAs_2$ NW cross-section showing lattice arrangement of **a** Tantalum (Ta), and **b** Arsenic (As) atoms.

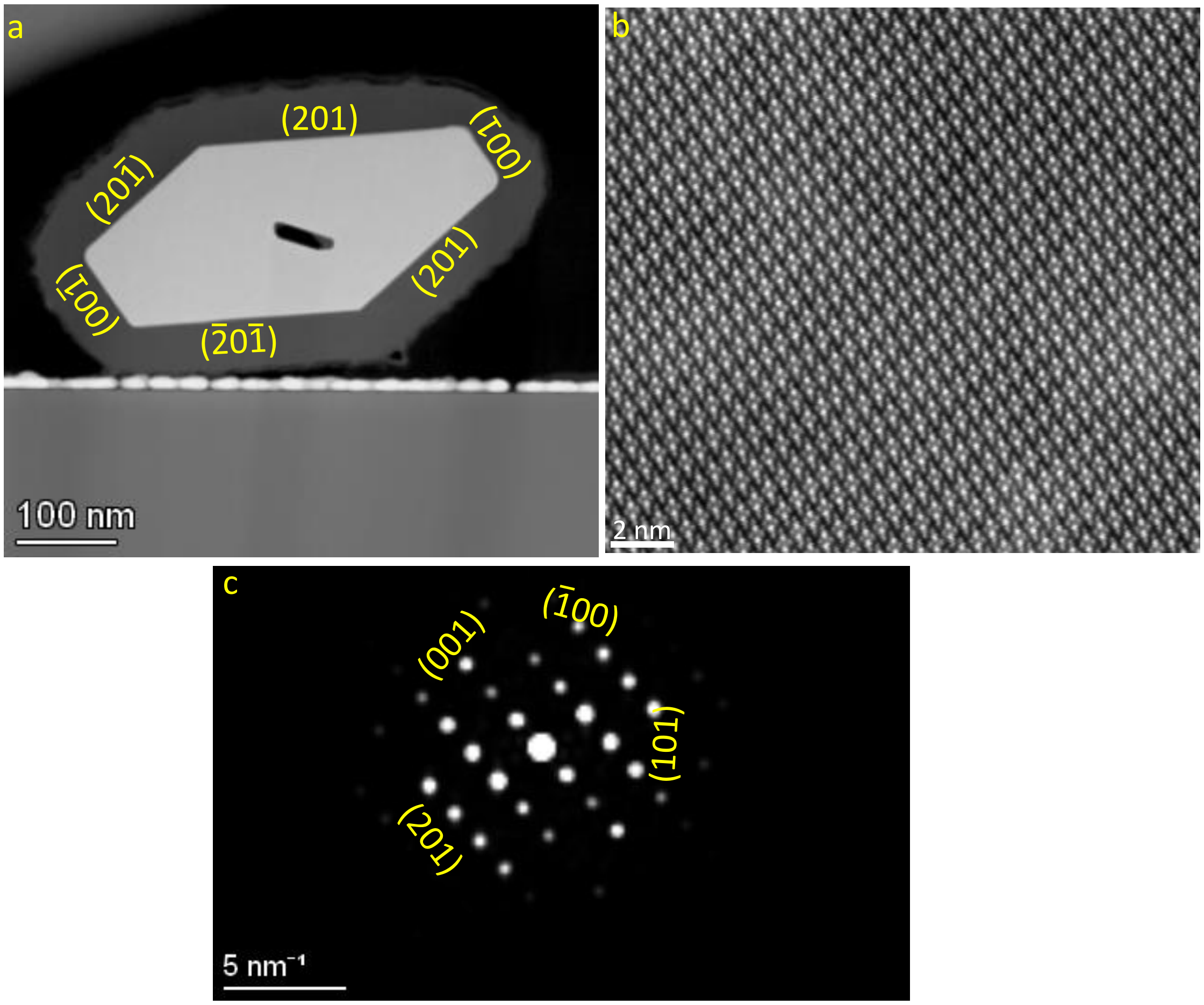

**FigS4.** **a** Low-magnification HAADF-STEM image of a NW cross-section showing $TaAs_2@SiO_2$ core-shell structure. **b** HAADF-STEM atomic resolution image from the core showing highly ordered $TaAs_2$ crystal. **c** FFT pattern from the center of **b** showing single-crystal nature of $TaAs_2$ with lattice planes.

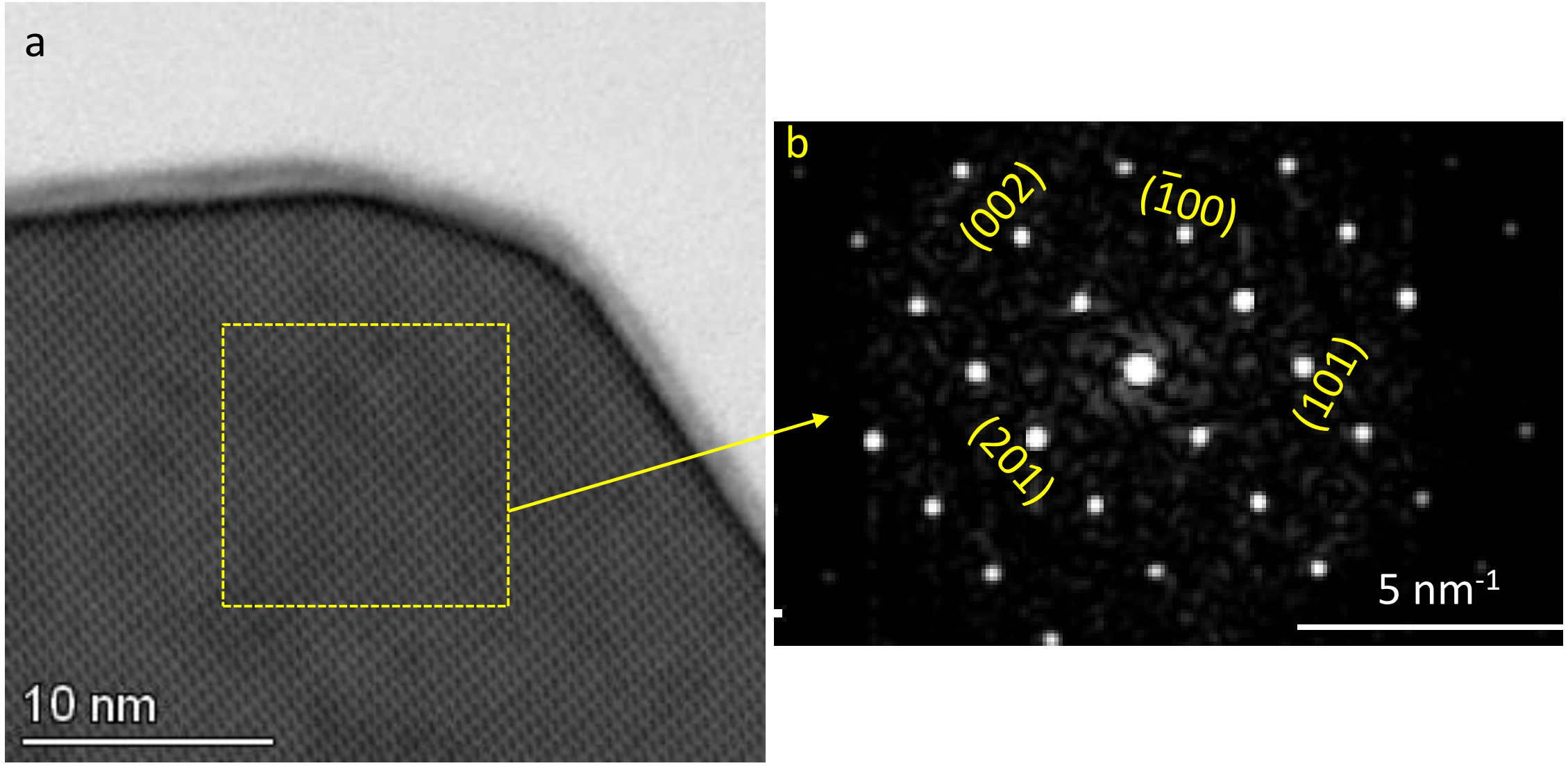

**FigS5. a** High-resolution BF-STEM image of interface region showing highly ordered $TaAs_2$ core and amorphous $SiO_2$-shell. **b** The FFT pattern from the square-box area exhibits single-crystalline features of the $TaAs_2$ surface.

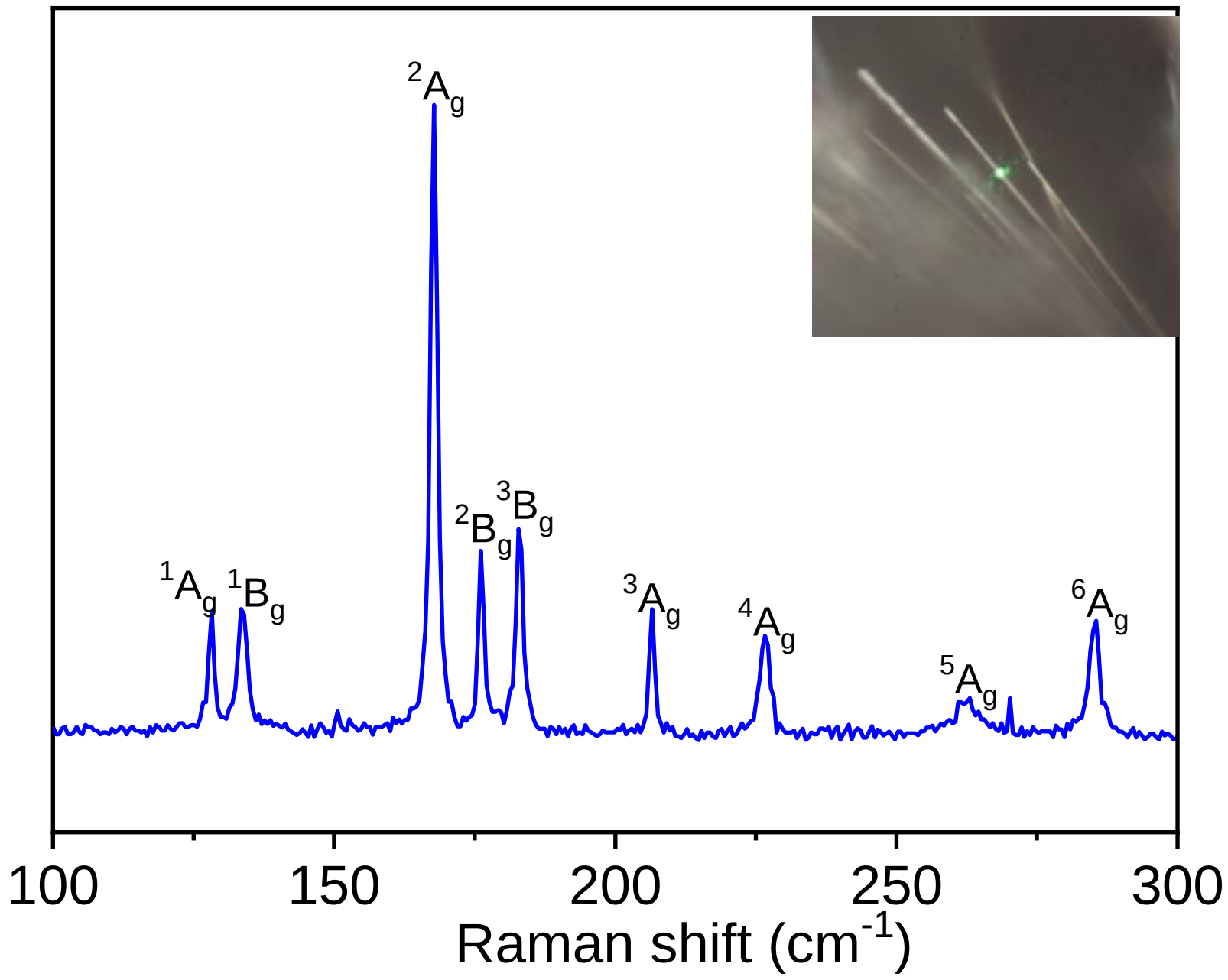

**Fig S6.** The Raman spectrum of $TaAs_2$ NW under a 532 nm (green) laser excitation assigned peaks agrees with the bulk $TaAs_2$. Inset: Optical Image of NW under laser excitation.

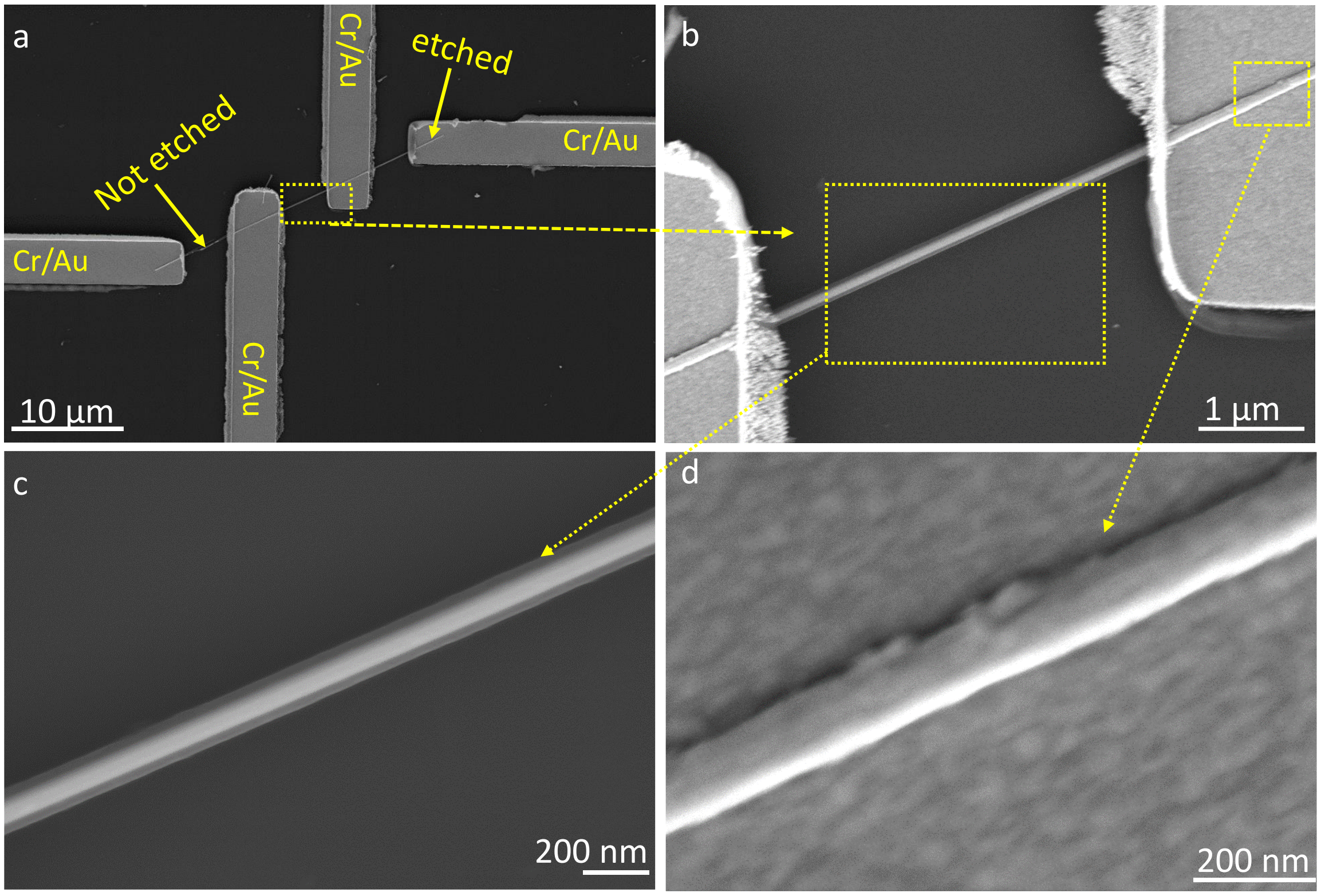

**Fig S7.** (a) SEM Image showing a typical four-terminal NW device. (b) low, and (c) high-magnification images showing non-etched part of NW shell between the electrodes (Cr/Au) wherein $SiO_2$ shell encapsulating $TaAs_2$ core can be seen, (b) low and (d) high magnification images showing part of NW wherein $SiO_2$ shell were selectively etched after lithography to make an electrical contact between $TaAs_2$-core and electrodes (Cr/Au:10/250 nm).

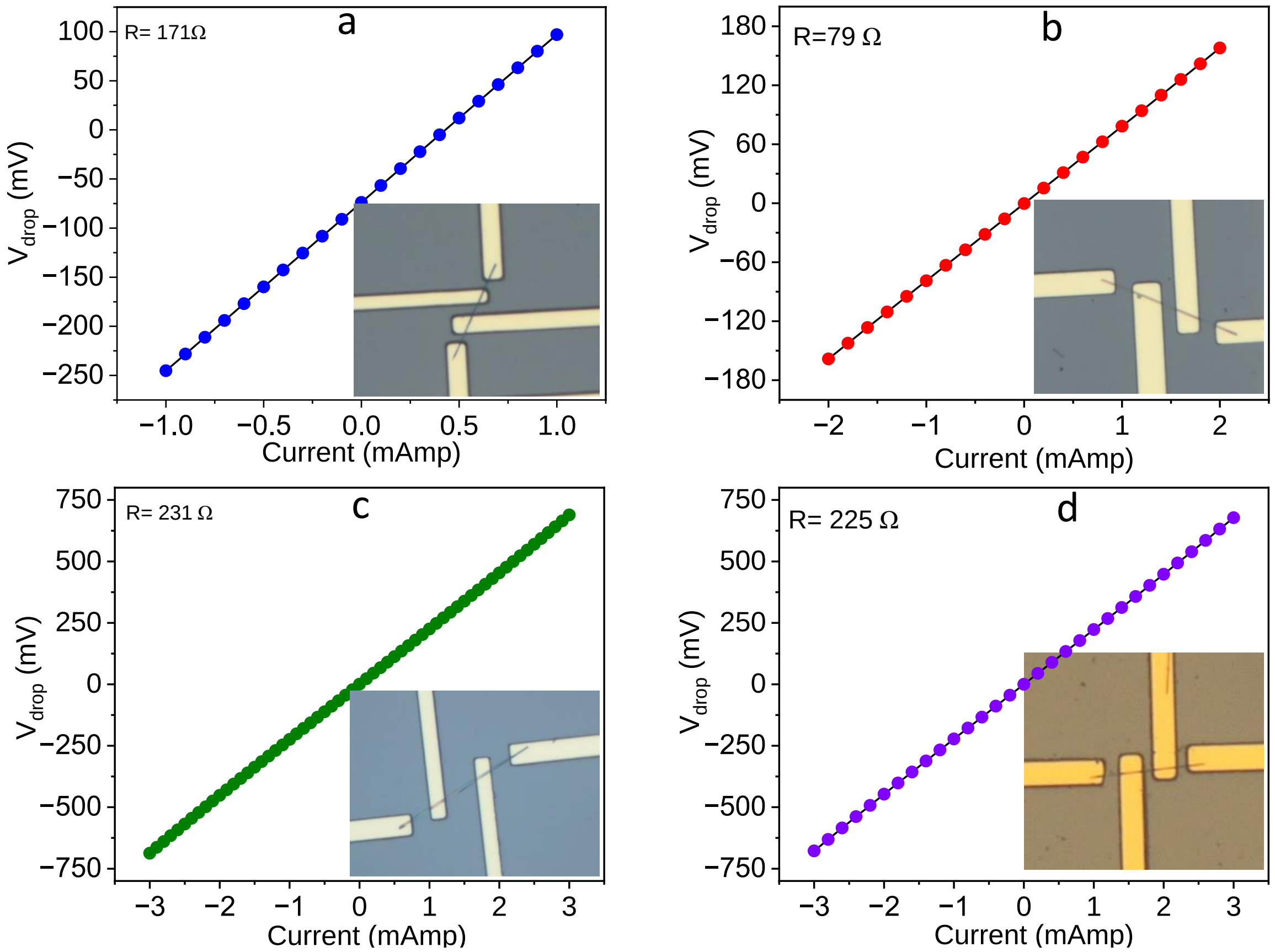

**Fig S8.** Ambient condition four-probe current-voltage (I-V) curves of different core diameter $TaAs_2$ NW devices under applied dc-currents (1 to 3 mA). Insets show optical images of NW-devices.

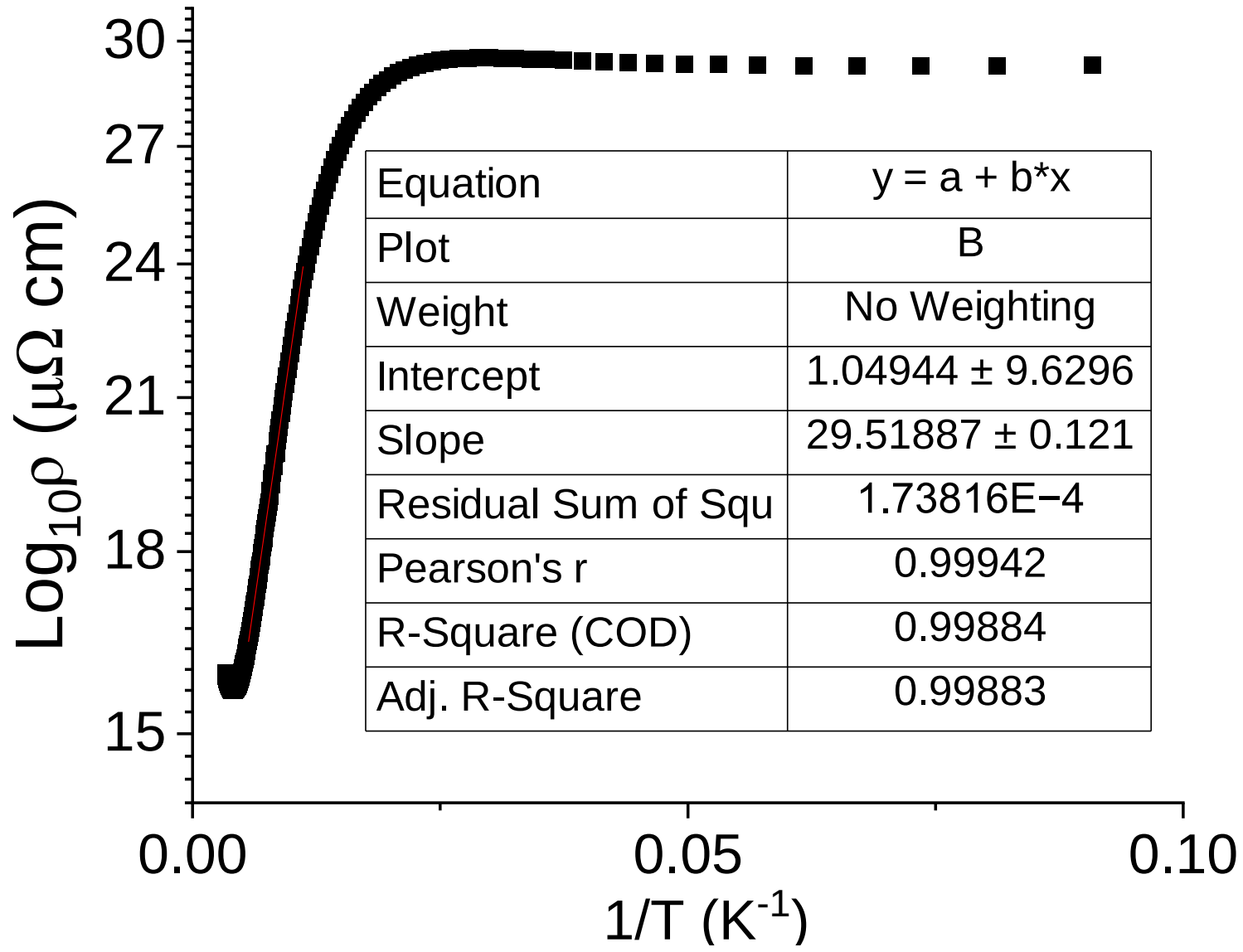

**Fig S9.** A linear fit of insulating state resistivity $\log_{10}\rho$ as a function of $T^{-1}$ employing Arrhenius equation for a 300 nm core diameter $TaAs_2$ NW.

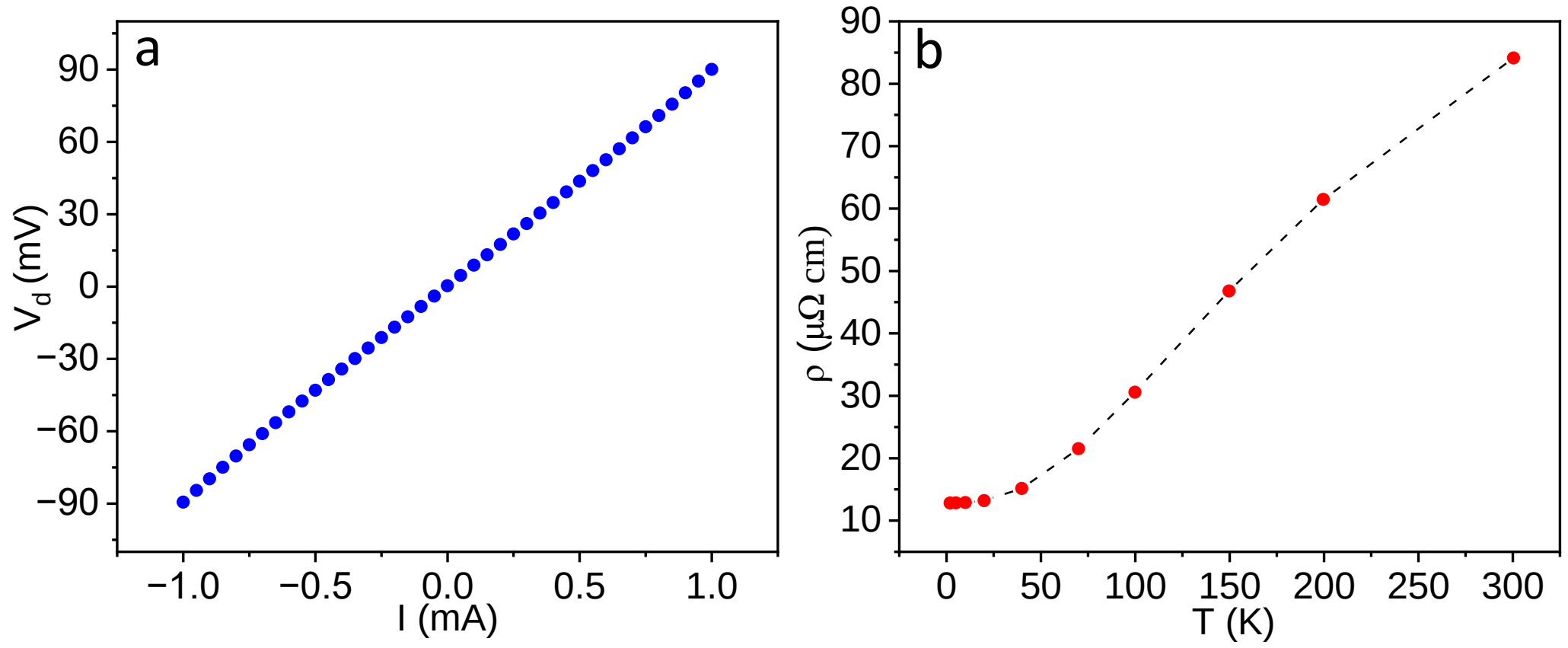

**Fig S10.** (a) Room-temperature I-V curve, (b) resistivity as function of temperature in a $TaAs_2$ NW with core diameter ~270 nm.

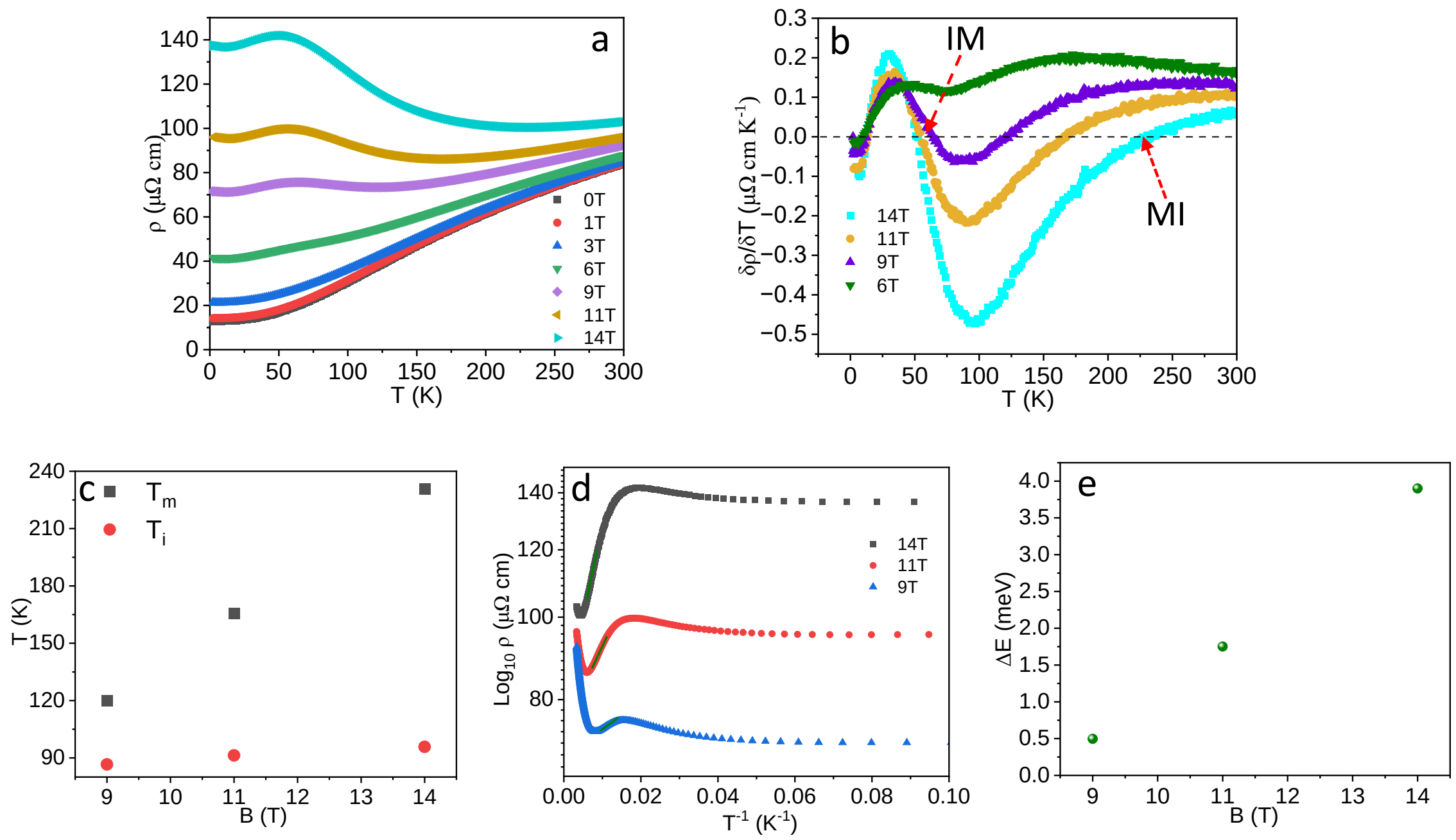

**Fig S11. Magnetotransport properties in a $TaAs_2$ NW device with ~ 270 nm core diameter**. **a** Resistivity as a function of temperature (RT) under increasing 0 to 14 T field. **b** The first derivative of resistivity $\delta\rho/\delta T$ as a function of temperature shows metal-to-insulator (MI) and novel insulator-to-metal (IM) transitions. **c** MI transition temperature ($T_m$) and inflection point temperature ($T_i$) as a function of increasing field strength. **d** $Log_{10}$ ($\rho$) versus $T^{-1}$ Arrhenius plots under varying field strengths with the fitting of the insulating regime to obtain gaps. **e** The obtained insulating gaps derived from **d** increase with magnetic-field strength.

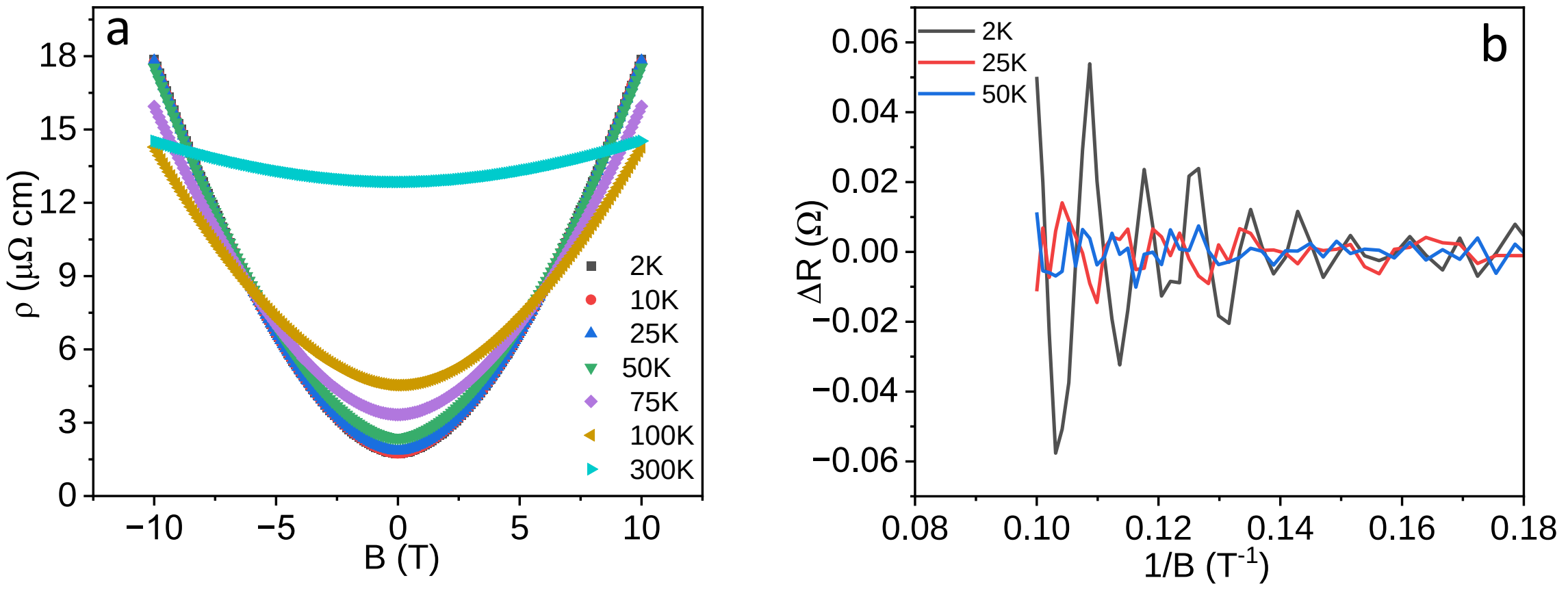

**Fig S12. Giant magnetoresistance (GMR) in ~ 270 nm $TaAs_2$ NW device. a** GMR as a function of temperature. GMR of NW manifest certain robust feature against temperature and decays much slower than the decay observed in bulk $TaAs_2$. **b** Shubnikov-de Hass (SdH)-oscillations as a function of 1/B under 2, 25, and 50 K.

**Effect of electrical gating on GMR**

In compensated semimetals, the magnitude of the MR is related to the degree of carrier compensation. For instance, in thin-$WTe_2$ flakes, MR was observed to peak at the charge neutrality point.[1] We have studied the effect of gate voltage ($V_g$) on the MR of our NW devices, employing two different approaches. In the first approach, the device was fabricated on Si/$SiO_2$ (300 nm $SiO_2$ with/without 20 nm ALD-$HfO_2$), and the gate dielectric was a thermal oxide layer on the silicon wafer wherein degenerately doped silicon acted as a back gate (**Fig.S13a**). In the second approach, we exploited the *in-situ* grown $SiO_2$ shell of NWs as the dielectric layer in a top-gate configuration (**Fig.S13c**). This approach is simpler and more efficient because the thickness of the $SiO_2$ shell is 60-80 nm, *i.e.*, ~ 4-5 times thinner than the thermal oxide; hence we expect a stronger field effect. In **Fig.S13a, S13c**, we present the effect of $V_g$ on the resistivity of NW devices at 1.8 K under a fixed field of 14 T. In the two device configurations, we observed an increase of MR under a negative to positive $V_g$ scan with an expected relatively faster response in the device with $SiO_2$-shell as top dielectric (**Fig. S13b, S13d**). The increase of MR with positive $V_g$ indicates that the NWs are slightly p-doped; thus, charge compensation occurs under positive $V_g$. The naturally occurring $SiO_2$ shell hence exhibits a double benefit: it acts as a removable protective layer for electrical contact and a dielectric layer for gating.

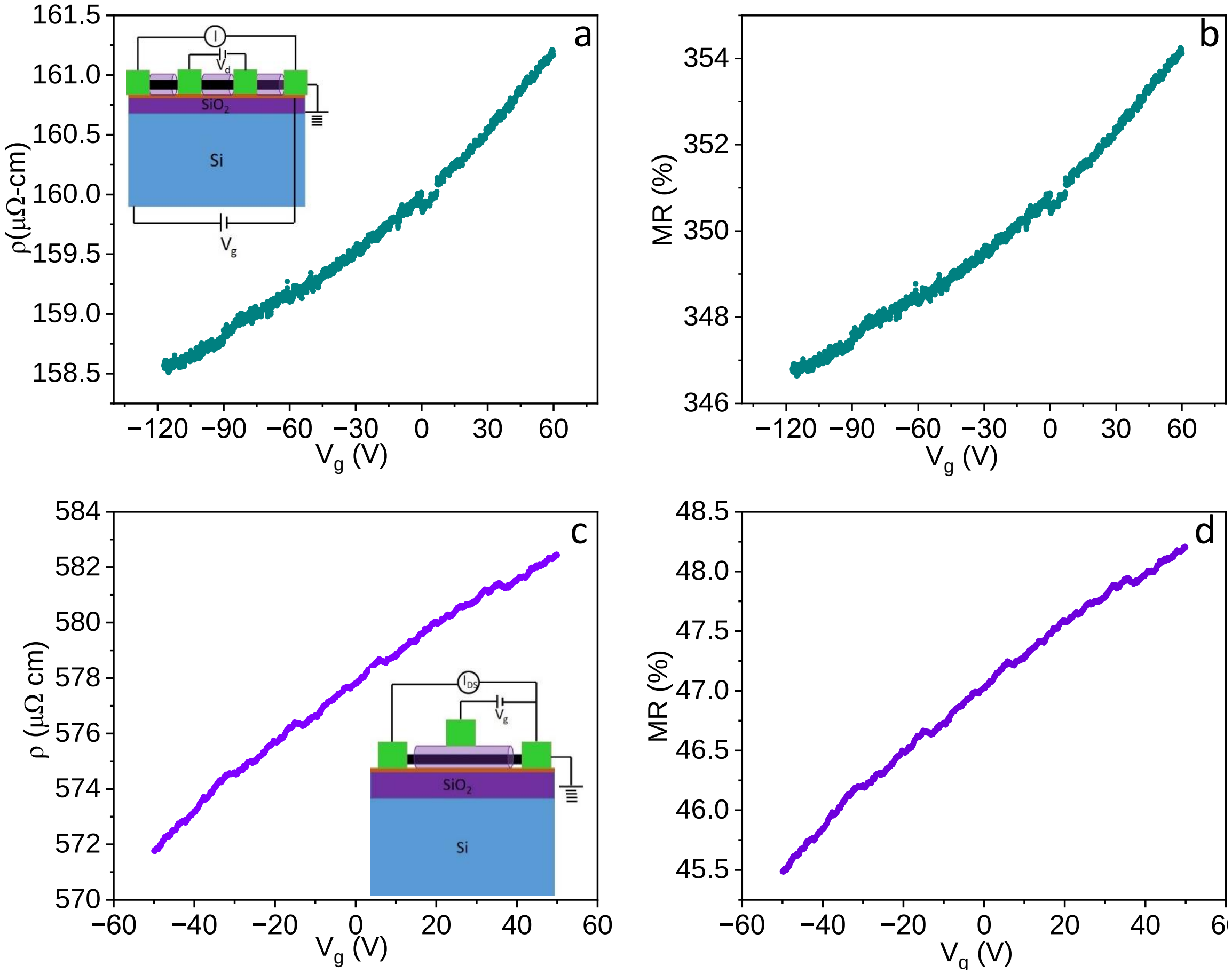

**Fig S13. Effect of Gate Voltage ($V_g$) on the MR of devices.** (a) Resistivity as a function of $V_g$ in a $TaAs_2$ NW (*d* ~ 125 nm) using Si back gate configuration. (b) Corresponding MR change as a function of $V_g$. (c) Resistivity as a function of $V_g$ in a $TaAs_2$ NW (*d* ~ 63 nm) using in-situ $SiO_2$-shell of NW as a top-dielectric. (d) Corresponding MR change as a function of $V_g$. Gate biasing was applied at a constant *B* of 14 T at 1.8 K.

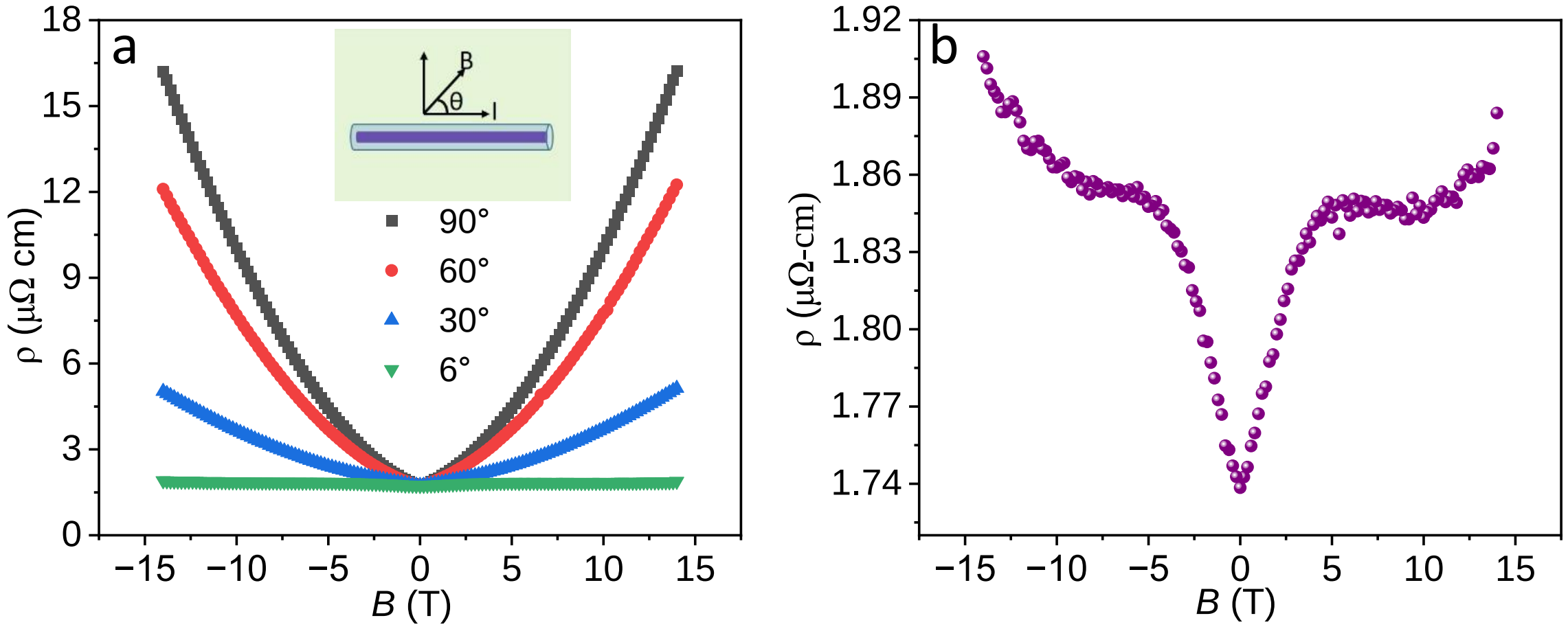

**Fig S14. Anisotropic GMR in a 300 nm core $TaAs_2$ NW-device. a** Magnetoresistance (MR) as a function of the scanned field under the different orientations of the NW-axis with respect to the field direction (*θ*). **b** Resistivity as a function of the scanned field at θ ~ 6° (close to longitudinal field configuration).

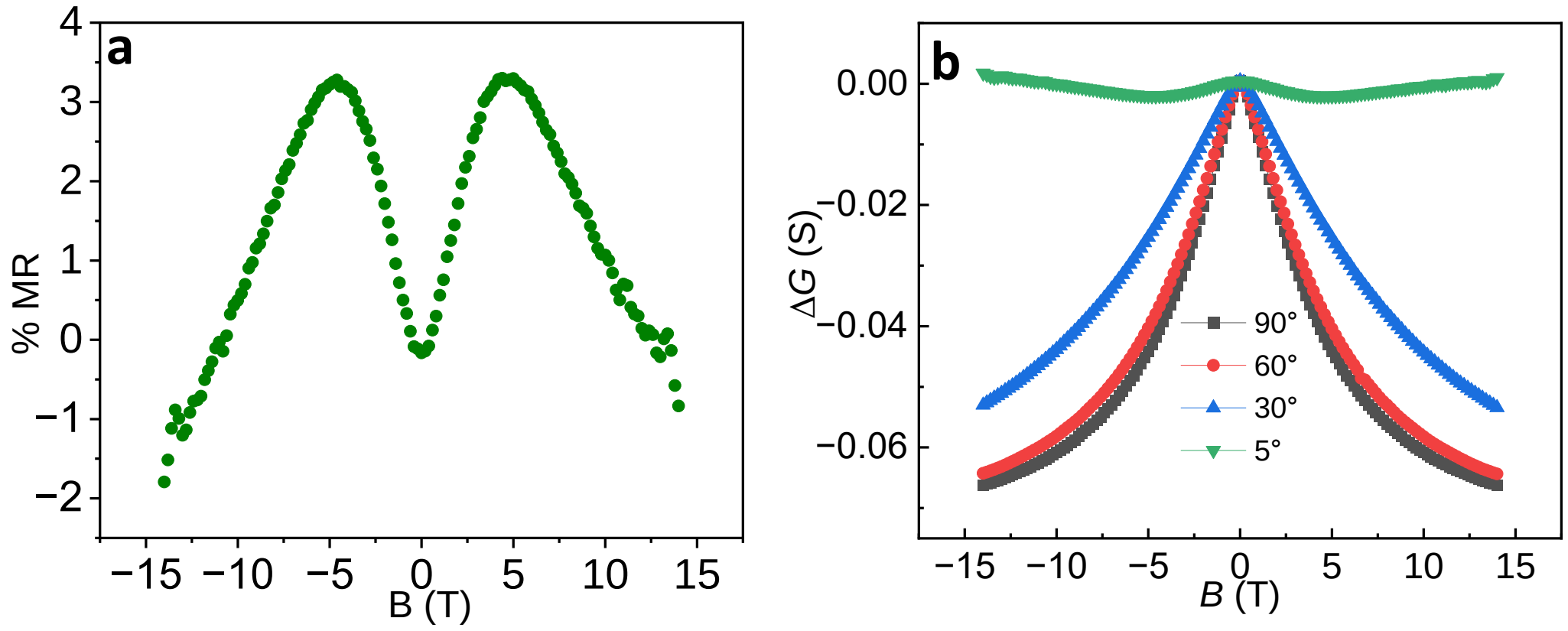

**Fig S15. a** A sharp dip (cusp) in MR near zero-field followed by small-field positive MR turning into negative MR at higher fields. **b** Change in magnetoconductance measured under varying *θ* from 90 to 6° for a *d* ~ 270 nm NW.

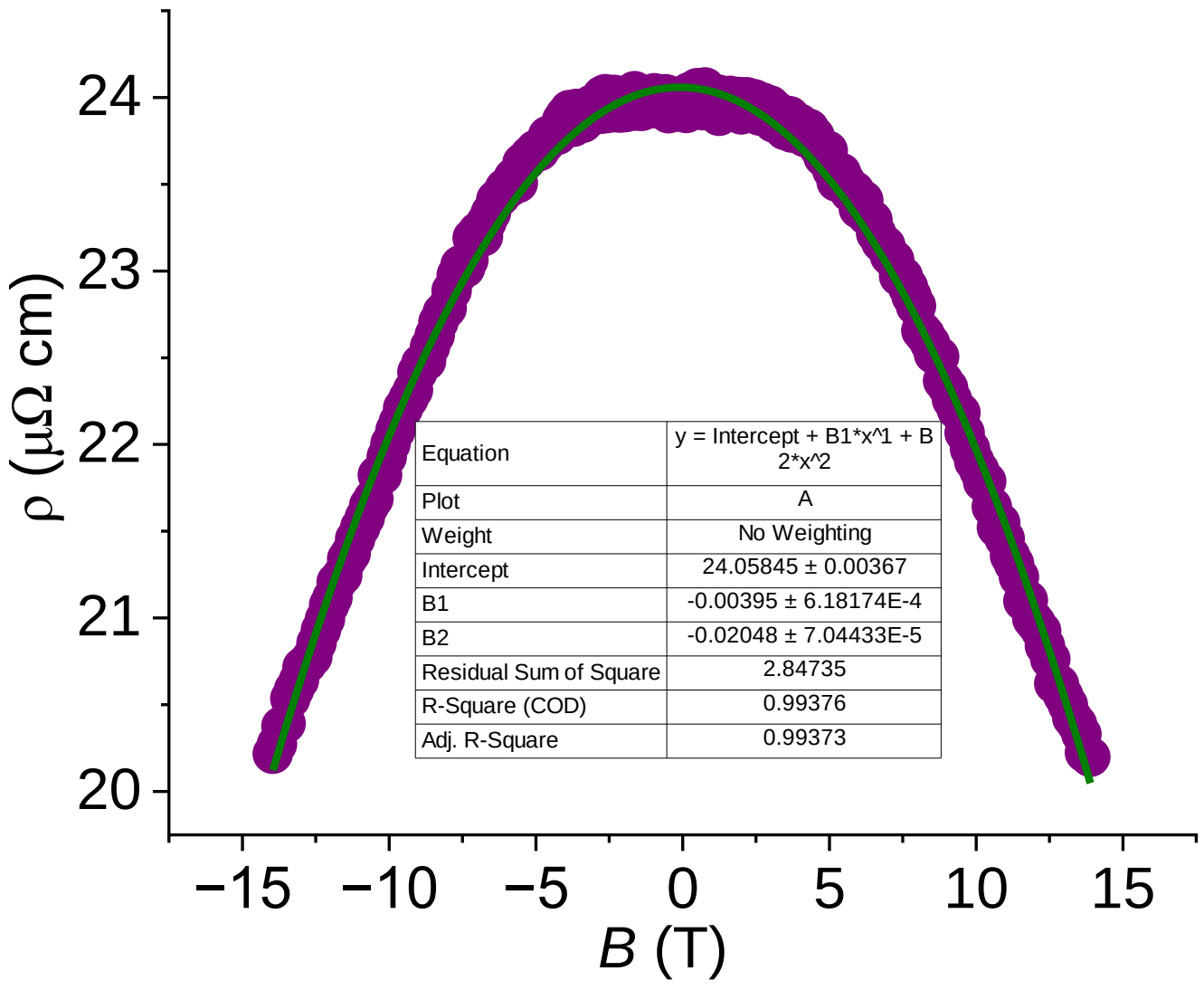

**Fig S16.** Fitting of field dependent longitudinal negative magnetoresistance curve for a 128 nm $TaAs_2$ NW-device at θ ~ 0°.

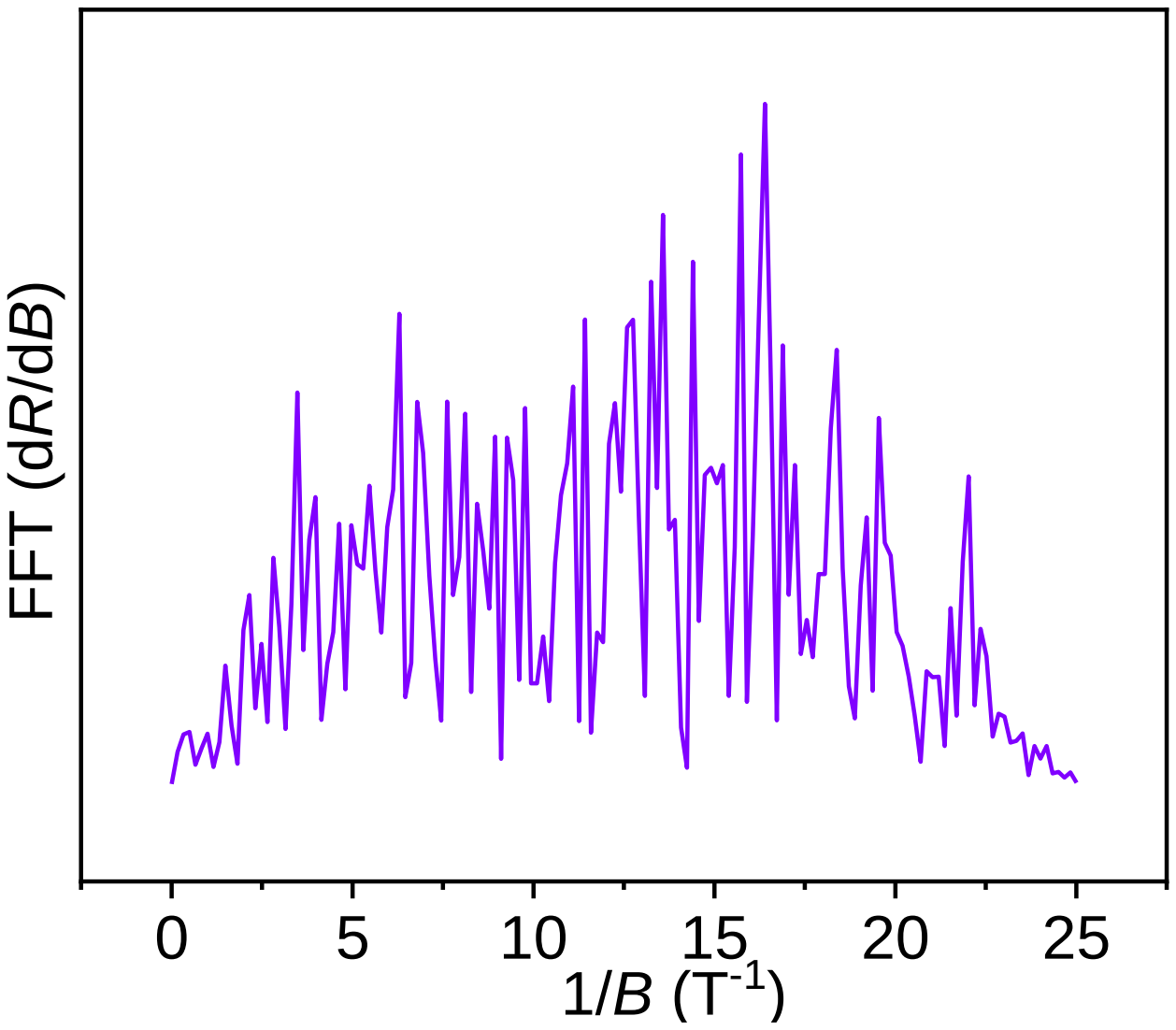

**Fig S17.** FFT spectrum of MR oscillations in a 128 nm $TaAs_2$ NW showing signatures of universal conductance fluctuation (UCF) and no apparent significant peak.

**AB-oscillations from the surface states of a WTI**

To gain intuition about the possible origin of a different AB oscillation pattern, we consider a simple model following Ref[2], of two Dirac cones on a two-dimensional surface BZ, away from the $\mathbf{\Gamma}$ point, for which the Dirac points are located at two surface momenta $(k_{x1}, k_{y1}), (k_{x2}, k_{y2})$ (**Figs. 4i and S18a**). For one of these cones, the low energy Hamiltonian is $\mathrm{H} = \hbar v_x(k_x - \mathrm{k_{x1}})\sigma_x + \hbar v_y(k_y - k_{y1})\sigma_y$, where $v_x, v_y$ represent the Fermi velocities in $x$ and $y$ directions, $\mathbf{k} = (k_x, k_y)$ is the surface momentum, and $\boldsymbol{\sigma}$ are a set of Pauli matrices. In a NW geometry, and under the assumption that the NW is cylindrical with a radius $R$ along the $y$-axis, $k_y$ remains a good quantum number, however, $k_x$ is now quantized to discrete angular momentum values, which take the form $l_n = n + \frac{1}{2}$ with $n$ an integer. Performing a coordinate transformation, the surface Dirac cones are now discretized to a set of bands that disperse along $k_y$ (Fig.1b), and are described by the energies:

$$E_{n,k_1}(k_y, \phi) = \pm\hbar\sqrt{\left[v_x^2\left(\frac{l_n + \phi}{R} - k_{x1}\right)^2 + v_y^2\left(k_y - k_{y1}\right)^2\right]}$$

The NW is threaded with magnetic field along its axis, which is included in the energy spectrum through a shift to the angular momentum: $\phi = \frac{\Phi}{\Phi_0}$ represent the number of magnetic flux quanta through the wire, where $\Phi_0 = \frac{h}{e}$ is a flux quantum and $\Phi = \mathbf{B} \cdot \mathbf{A}$ is the total flux threaded by magnetic field $\mathbf{B}$ through the wire's cross section $\mathbf{A}$.

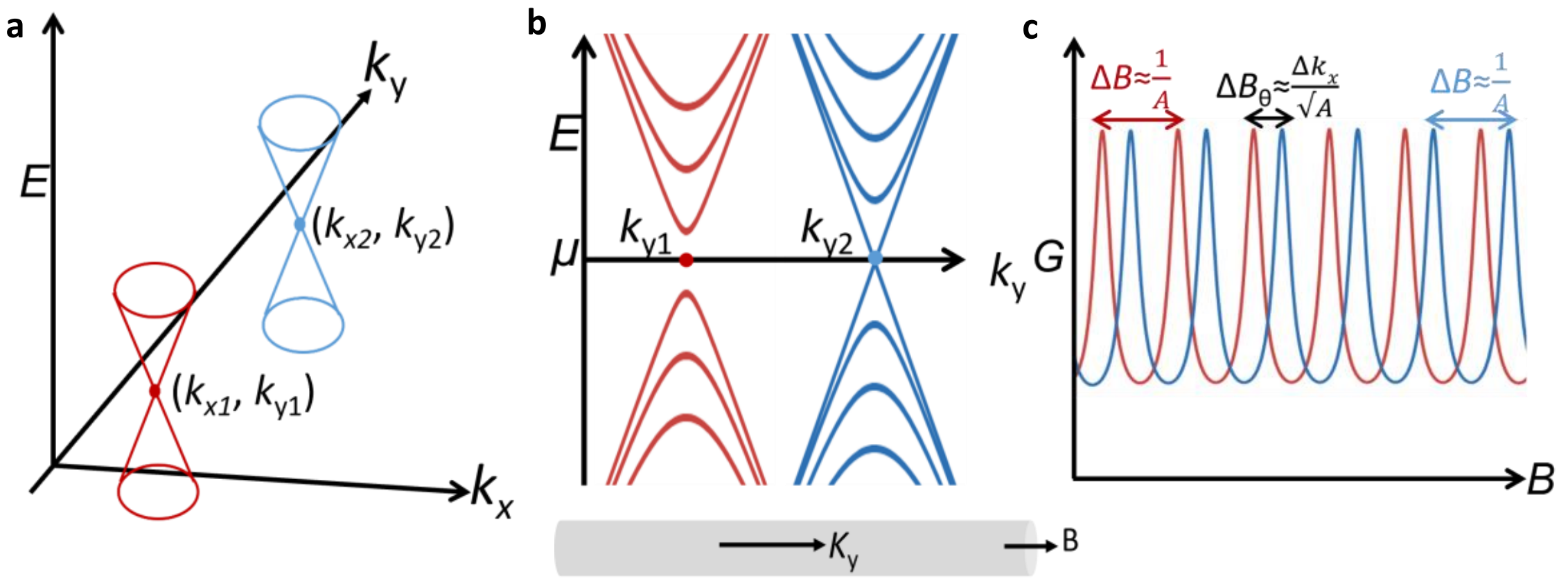

**Fig S18.** Model of AB oscillations of two surface Dirac cones. **a** Surface spectrum with two Dirac cones located at $(k_{x1}, k_{y1}), (k_{x2}, k_{y2})$. **b** In the NW geometry, the surface Dirac cones become two sets of discrete energy bands, and the magnetic field B shifts them in energy. For general B values the spectrum is gapped, however, for some B values the gap of one or more of the Dirac cones is closed and the bands cross the chemical potential. **c** Each Dirac cone will show oscillations in the conductance G as a function of B, with the same period $\Delta B \sim \frac{1}{A}$ (presented in red and blue). However, between the two periods a phase shift can appear $\Delta B_\theta \sim \frac{\Delta k_x}{\sqrt{A}}$. The total conductance (black) can then get a complex oscillations pattern.

Tuning the magnetic field will change the discrete bands presented (**Fig.S18b**), shifting them in energy. Assuming that the chemical potential is located at $E = 0$, the spectrum is generally gapped, however, for some magnetic field values, it is gapless with a linear-dispersing mode crossing the chemical potential. Therefore, the conductance will show oscillations with increasing magnetic fields. (For other chemical potential values, these oscillations will appear due to the change in the of the number of modes[3]. For each Dirac cone, the values in which the spectrum is gapless, namely $E_{n,k_1}(k_y, \phi) = 0$, will appear for momentum $k_y = k_{y1}$ and for magnetic field values

$$\boldsymbol{B_{n,k_1}} = \frac{\Phi_0}{\mathbf{A}}(k_{x1}R - l_n)$$

The period of the oscillations will be the difference between two values of such magnetic fields:

$$\Delta\mathbf{B} = B_{n,k_1} - B_{n+1,k_1} = \frac{\Phi_0}{\mathbf{A}}$$

The period of oscillations for each Dirac cone is therefore inversely proportional to the cross-section, **A**. In the presence of more than one Dirac cone, another oscillation pattern will appear, with the same period (as $\Delta B$ is independent of $k_{x1}$). However, between the two patterns there may be a phase shift:

$$\Delta\mathbf{B}_\theta = \mathrm{B}_{n,k_1} - \mathrm{B}_{n,k_2} = \frac{\Phi_0\sqrt{\pi}}{\sqrt{\mathbf{A}}}\Delta k_x$$

With $\Delta k_x = (k_{x1} - k_{x2})$. Since the total conductance will have contributions from both Dirac cones, it will show a complex AB oscillation pattern, as illustrated in **Fig.S18c**.

We stress that the actual transport measurements will be influenced by many factors not included in this calculation, such as the existence and number of Dirac cones/Fermi arcs in the wire's surfaces' spectra, their location in the BZ, their energy relative to the chemical potential/bulk gap, and the way they connect between the different facets of the NW. However, the model presented here demonstrates how in the presence of a complex surface states spectrum, beyond a single Dirac cone at the **Γ** point, a richer pattern of AB oscillations may appear. Specifically, the double pattern presented here could be interpreted as a pattern with a halved oscillation period, and hence an effective wire cross-section twice larger than the real cross-section, as we see in our results.

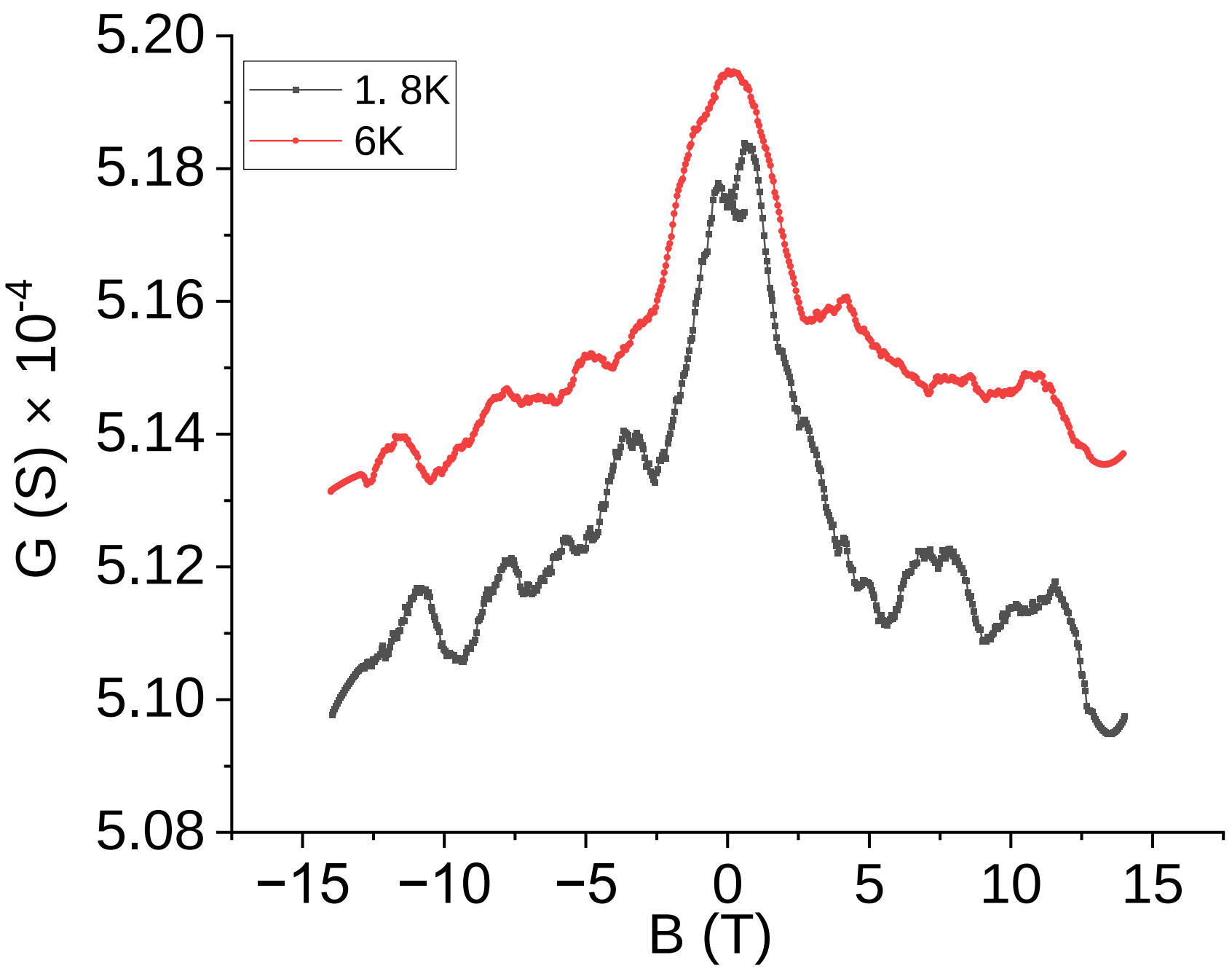

**Fig S19.** Magnetoconductance at 2 and 6 K as a function of magnetic field for *d*~ 32 nm NW device.